\pdfoutput=1

\documentclass[letterpaper,11pt]{article} 

\usepackage{fullpage}
\usepackage{graphicx}
\usepackage{amssymb}
\usepackage{amsmath}
\usepackage{amsthm} %removed this on 12th April because of latex error.
\usepackage{natbib}
\usepackage{hyperref}
\usepackage{afterpage}
\usepackage{rotating}

\newcommand{\bfzero}{\mathbf{0}}

\newcommand{\bfA}{\mathbf{A}}
\newcommand{\bfB}{\mathbf{B}}
\newcommand{\bfC}{\mathbf{C}}
\newcommand{\bfD}{\mathbf{D}}

\newcommand{\bfI}{\mathbf{I}}

\newcommand{\bfM}{\mathbf{M}}

\newcommand{\bfU}{\mathbf{U}}
\newcommand{\bfV}{\mathbf{V}}
\newcommand{\bfW}{\mathbf{W}}
\newcommand{\bfX}{\mathbf{X}}
\newcommand{\bfY}{\mathbf{Y}}
\newcommand{\bfZ}{\mathbf{Z}}

\newcommand{\bbR}{\mathbb{R}}
\newcommand{\bbX}{\mathbb{X}}

\newcommand{\mcalL}{\mathcal{L}}

\newcommand{\cov}{\mbox{cov}}

\newcommand{\trace}{\mbox{tr}}
\newcommand{\diag}{\mbox{diag}}

\newcommand{\figurewidth}{1.00}
\newcommand{\raiselegendtext}{1.0cm}
\newcommand{\norml}{\ell_}
\newcommand{\aicc}{\mbox{AIC}_{\mbox{c}}}

\newtheorem{theorem}{Theorem}
\newtheorem{lemma}{Lemma}

\begin{document}

\title{A path following algorithm for \\
Sparse Pseudo-Likelihood Inverse Covariance Estimation \\
(SPLICE)}
\author{Guilherme V. Rocha, Peng Zhao, Bin Yu}
\maketitle

%\singlespacing

\begin{abstract}
	Given $n$ observations of a $p$-dimensional random vector, the covariance matrix and its inverse (precision matrix) are needed in a wide range of applications.
	Sample covariance (e.g. its eigenstructure) can misbehave when $p$ is comparable to the sample size $n$.
	Regularization is often used to mitigate the problem.

	In this paper, we proposed an $\ell_1$ penalized pseudo-likelihood estimate for the inverse covariance matrix. 
	This estimate is sparse due to the $\ell_1$ penalty, and we term this method SPLICE.
	Its regularization path can be computed via an algorithm based on the homotopy/LARS-Lasso algorithm. 
	Simulation studies are carried out for various inverse covariance structures for $p=15$ and $n=20,1000$.
	We compare SPLICE with the $\ell_1$ penalized likelihood estimate and a $\ell_1$ penalized Cholesky decomposition based method.
	SPLICE gives the best overall performance in terms of three metrics on the precision matrix and ROC curve for model selection. 
	Moreover, our simulation results demonstrate that the SPLICE estimates are positive-definite for most of the regularization path even though the restriction is not enforced.		
\end{abstract}

%\singlespacing

%%%%%%%%%%%%%%%%%%%%%%%%%%%%%%%%%%%%%%%%%%%%%%%%%%%%%%%%%%%%%%%%%%%%%%%%%%%%%%%%%%%%%%%%%%%%%%%%%%%%%%%%%%

\section*{Acknowledgments}

The authors gratefully acknowledge the support 
of NSF grant 
DMS-0605165, ARO grant W911NF-05-1-0104, NSFC (60628102), and a grant from MSRA.
B. Yu also thanks the Miller Research Professorship 
in Spring 2004 from the Miller Institute at University of California at Berkeley and a 2006 Guggenheim Fellowship. 
G. Rocha also acknowledges helpful comments by Ram Rajagopal, Garvesh Raskutti, Pradeep Ravikumar and Vincent Vu.

%%%%%%%%%%%%%%%%%%%%%%%%%%%%%%%%%%%%%%%%%%%%%%%%%%%%%%%%%%%%%%%%%%%%%%%%%%%%%%%%%%%%%%%%%%%%%%%%%%%%%%%%%%

\section{Introduction}
\label{section:splice_introduction}

Covariance matrices are perhaps the simplest statistical measure of association between a set of variables and widely used.
Still, the estimation of covariance matrices is extremely data hungry, as the number of fitted parameters grows rapidly with the number of observed variables $p$.
Global properties of the estimated covariance matrix, such as its eigenstructure, are often used \citep[e.g. Principal Component Analysis, ][]{jolliffe:2002:principal-component-analysis}.
Such global parameters may fail to be consistently estimated when the number of variables $p$ is non-negligible in the comparison to the sample size $n$.
As one example, it is a well-known fact that the eigenvalues and eigenvectors of an estimated covariance matrix are inconsistent when the ratio $\frac{p}{n}$ does not vanish asymptotically
\citep{marchenko:1967:the-distribution-of-eigenvalues-in-certain-sets-of-random-matrices,paul:2007:pre-conditioning-for-feature-selection-and-regression-in-high-dimensional-problems}.
Data sets with a large number of observed variables $p$ and small number of observations $n$ are now a common occurrence in statistics.
Modeling such data sets creates a need for regularization procedures capable of imposing sensible structure on the estimated covariance matrix while being computationally efficient.

Many alternative approaches exist for improving the properties of covariance matrix estimates.
\textit{Shrinkage methods} for covariance estimation were first considered in \citet{stein:1975:estimating-a-covariance-matrix, stein:1986:lectures-on-the-theory-of-estimation-of-many-parameters} as a way to correct the overdispersion of the eigenvalues of estimates of large covariance matrices.
\citet{ledoit:2004:large_covariance_matrix} present a shrinkage estimator that is 
the asymptotically optimal convex linear combination of the sample covariance matrix and the identity matrix with respect to the Froebenius norm.
\citet{daniels:1999:nonconjugate-bayesian-estimation-of-covariance-matrices-and-its-use-in-hierarchical-models,daniels:2001:shrinkage-estimators-for-covariance-matrices} propose alternative strategies using shrinkage toward diagonal and more general matrices.
\textit{Factorial models} have also been used as a strategy to regularize estimates of covariance matrices \citep{fan:2006:high-dimensional-covariance-matrix-estimation-using-a-factor-model}.
\textit{Tapering} the covariance matrix is frequently used in time series and spatial models and have been used recently to improve the performance of covariance matrix estimates used by classifiers based on linear discriminant analysis \citep{bickel:2004:some-theory-for-fishers-linear-discriminant-function-naive-bayes-and-some-alternatives} and in Kalman filter ensembles \citep{furrer:2007:estimation-of-high-dimensional-prior-and-posterior-covariance-matrices-in-kalman-filter-variants}.
Regularization of the covariance matrix can also be achieved by \textit{regularizing its eigenvectors} \citep{johnstone:2004:sparse-principal-component-analysis, zou:2004:sparse_pca}.

\textit{Covariance selection} methods for estimating covariance matrices consist of imposing sparsity on the precision matrix (i.e., the inverse of the covariance matrix).
The Sparse Pseudo-Likelihood Inverse Covariance Estimates (SPLICE) proposed in this paper fall into this category.
This family of methods was introduced by \citet{dempster:1972:covariance-selection}.
An advantage of imposing structure on the precision matrix stems from its close connections to linear regression.
For instance, \citet{wu:2003:nonparametric-estimation-of-large-covariance-matrices-of-longitudinal-data} use, for a fixed order of the random vector, a parametrization of the precision matrix $\bfC$ in terms of a decomposition $\bfC = \bfU'\bfD\bfU$ with $\bfU$ upper-triangular with unit diagonal and $\bfD$ a diagonal matrix.
The parameters $\bfU$ and $\bfD$ are then estimated through $p$ linear regressions and Akaike's Information Criterion \citep[AIC, ][]{akaike:1973:information-criterion} is used to promote sparsity in $\bfU$.
A similar covariance selection method is presented in \citet{bilmes:2000:factored-sparse-inverse-covariance-matrices}.
More recently, \citet{bickel:2006:regularized-estimation-of-large-covariance-matrices} have obtained conditions ensuring consistency in the operator norm (spectral norm) for precision matrix estimates based on banded Cholesky factors.
Two disadvantages of imposing the sparsity in the factor $\bfU$ are: sparsity in $\bfU$ does not necessarily translates into sparsity of $\bfC$ and; the sparsity structure in $\bfU$ is sensitive to the order of the random variables within the random vector.
The SPLICE estimates proposed in this paper constitute an attempt at tackling these issues.

The AIC selection criterion used in \citet{wu:2003:nonparametric-estimation-of-large-covariance-matrices-of-longitudinal-data}
requires, in its exact form, that the estimates be computed for all subsets of the parameters in $\bfU$.
A more computationally tractable alternative for performing parameter selection consists in penalizing parameter estimates by their $\norml{1}$-norm \citep{breiman:1995:nonnegative_garrote, tibshirani:1996:lasso, chen:2001:basis-pursuit}, popularly known as the LASSO in the context of least squares linear regression.
The computational advantage of the $\norml{1}$-penalization over penalization by the dimension of the parameter being fitted ($\norml{0}$-norm) -- such as in AIC -- stems from its convexity \citep{boyd:2004:convex_optimization}.
Homotopy algorithms for tracing the entire LASSO regularization path have recently become available \citep{osborne:2000:lasso_dual, efron:2004:lars}.
Given the high-dimensionality of modern days data sets, it is no surprise that $\norml{1}$-penalization has found its way into the covariance selection literature.

\citet{huang:2006:covariance-matrix-selection-and-estimation-via-penalized-normal-likelihood} propose a covariance selection estimate corresponding to an $\norml{1}$-penalty version of the Cholesky estimate of \citet{wu:2003:nonparametric-estimation-of-large-covariance-matrices-of-longitudinal-data}.
The off-diagonal terms  of $\bfU$ are penalized by their $\norml{1}$-norm and cross-validation is used to select a suitable regularization parameter.
While this method is very computationally tractable (an algorithm based on the homotopy algorithm for linear regressions is detailed below in Appendix \ref{appendix:cholesky_path_tracing}), it still suffer from the deficiencies of Cholesky-based methods.
Alternatively, \citet{banerjee:2005:sparse-covariance-selection-via-robust-maximum-likelihood-estimation}, \citet{banerjee:2007:model-selection-through-sparse-maximum-likelihood-estimation-for-multivariate-gaussian-or-binary},  \citet{yuan:2007:model-selection-and-estimation-in-the-gaussian-graphical-model}, and \citet{friedman:2008:sparse-inverse-covariance-estimation-with-the-graphical-lasso}
consider an estimate defined by the Maximum Likelihood of the precision matrix for the Gaussian case penalized by the $\norml{1}$-norm of its off-diagonal terms.
While these methods impose sparsity directly in the precision matrix, no path-following algorithms are currently available for them.
\citet{rothman:2007:sparse-permutation-invariant-covariance-estimation} analyze the properties of estimates defined in terms of $\norml{1}$-penalization of the exact Gaussian neg-loglikelihood and introduce a permutation invariant method based on the Cholesky decomposition to avoid the computational cost of semi-definite programming.

%This paper presents Sparse Pseudo-Likelihood Inverse Covariance Estimates (SPLICE)  as an alternative that imposes sparsity constraints directly on the precision matrix and for which homotopy algorithms can be used to trace the regularization path.
The SPLICE estimates presented here impose sparsity constraints directly on the precision matrix.
Moreover the entire regularization path of SPLICE estimates can be computed by homotopy algorithms.
It is based on previous work by \citet{meinshausen:2006:high-dimensional-graphs-and-variable-selection-with-the-lasso} for neighborhood selection in Gaussian graphical models.
While \citet{meinshausen:2006:high-dimensional-graphs-and-variable-selection-with-the-lasso} use $p$ separate linear regressions to estimate the neighborhood of one node at a time, we propose merging all $p$ linear regressions into a single least squares problem where the observations associated to each regression are weighted according to their conditional variances.
The loss function thus formed can be interpreted as a pseudo neg-loglikelihood \citep{besag:1974:spatial-interaction-and-the-statistical-analysis-of-lattice-systems} in the Gaussian case.
To this pseudo-negloglikelihood minimization, we add symmetry constraints and a weighted version of the $\norml{1}$-penalty on off-diagonal terms to promote sparsity.
The SPLICE estimate can be interpreted as an approximate solution following from replacing the exact neg-loglikelihood in \citet{banerjee:2007:model-selection-through-sparse-maximum-likelihood-estimation-for-multivariate-gaussian-or-binary} by a quadratic surrogate (the pseudo neg-loglikelihood).

The main advantage of SPLICE estimates is algorithmic: by use of a proper parametrization, the problem involved in tracing the SPLICE regularization path can be recast as a linear regression problem and thus amenable to be solved by a homotopy algorithm as in  \citet{osborne:2000:lasso_dual} and \citet{efron:2004:lars}.
To avoid computationally expensive cross-validation, we use information criteria to select a proper amount of regularization.
We compare the use of Akaike's Information criterion \citep[AIC, ][]{akaike:1973:information-criterion}, a small-sample corrected version of the AIC \citep[$\aicc$,][]{hurvich:1998:smoothing-parameter-selection-in-nonparametric-regression-using-an-improved-akaike-information-criterion} and the Bayesian Information Criterion \citep[BIC, ][]{schwartz:1978:bic} for selecting the proper amount of regularization.

We use simulations to compare SPLICE estimates to the $\norml{1}$-penalized maximum likelihood estimates \citep{banerjee:2005:sparse-covariance-selection-via-robust-maximum-likelihood-estimation, banerjee:2007:model-selection-through-sparse-maximum-likelihood-estimation-for-multivariate-gaussian-or-binary, yuan:2007:model-selection-and-estimation-in-the-gaussian-graphical-model, friedman:2008:sparse-inverse-covariance-estimation-with-the-graphical-lasso} and to the $\norml{1}$-penalized Cholesky approach in \citet{huang:2006:covariance-matrix-selection-and-estimation-via-penalized-normal-likelihood}.
We have simulated both small and large sample data sets.
Our simulations include model structures commonly used in the literature (ring and star topologies, AR processes) as well as a few randomly generated model structures.
SPLICE had the best performance in terms of the quadratic loss and the spectral norm of the precision matrix deviation ($\|\bfC-\hat\bfC\|_{2}$). 
It also performed well in terms of the entropy loss.
SPLICE had a remarkably good performance in terms of selecting the off-diagonal terms of the precision matrix: in the comparison with Cholesky, SPLICE incurred a smaller number of false positives to select a given number of true positives; in the comparison with the penalized exact maximum likelihood estimates, the path following algorithm allows for a more careful exploration of the space of alternative models.

The remainder of this paper is organized as follows.
Section \ref{section:pseudolikelihood_loss_function} presents our pseudo-likelihood surrogate function and some of its properties.
Section \ref{section:pseudo_likelihood_algorithm} presents the homotopy algorithm used to trace the SPLICE regularization path.
Section \ref{section:simulation_results_splice} presents simulation results comparing the SPLICE estimates with some alternative regularized methods.
Finally, Section \ref{section:discussion} concludes with a short discussion.

%%%%%%%%%%%%%%%%%%%%%%%%%%%%%%%%%%%%%%%%%%%%%%%%%%%%%%%%%%%%%%%%%%%%%%%%%%%%%%%%%%%%%%%%%%%%%%%%%%%%%%%%%%

\section{An approximate loss function for inverse covariance estimation}
\label{section:pseudolikelihood_loss_function}

In this section, we establish a parametrization of the precision matrix $\Sigma^{-1}$ of a random vector $\bbX$ in terms of the coefficients in the linear regressions among its components.
We emphasize that the parametrization we use differs from the one previously used by \citet{wu:2003:nonparametric-estimation-of-large-covariance-matrices-of-longitudinal-data}.
Our alternative parametrization is used to extend the approach used by  \citet{meinshausen:2006:high-dimensional-graphs-and-variable-selection-with-the-lasso} for the purpose of estimation of sparse precision matrices.
The resulting loss function can be interpreted as a pseudo-likelihood function in the Gaussian case.
For non-Gaussian data, the minimizer of the empirical risk function based on the loss function we propose still yields consistent estimates.
The loss function we propose also has close connections to linear regression and lends itself well for a homotopy algorithm in the spirit of \citet{osborne:2000:lasso_dual} and \citet{efron:2004:lars}.
A comparison of this approximate loss function to its exact counterpart in the Gaussian case suggests that the approximation is better the sparser the precision matrix.

In what follows, $\bfX$ is a $\bbR^{n\times p}$ matrix containing in each of its $n$ rows observations of the zero-mean random vector $\bbX$ with covariance matrix $\Sigma$.
Denote by $\bbX_{j}$ the $j$-th entry of $\bbX$ and by $\bbX_{J^{*}}$ the $(p-1)$ dimensional vector resulting from deleting $\bbX_{j}$ from $\bbX$.
For a given $j$, we can permute the order of the variables in $\bbX$ and partition $\Sigma$ to get:
\begin{eqnarray*}
\cov\left(
\left[
\begin{array}{c}
\bbX_{j} \\
\bbX_{J^{*}}
\end{array}
\right]
\right)
& = & 
\left[
\begin{array}{cc}
\sigma_{j,j}	& \Sigma_{j,J^{*}} \\
\Sigma_{J^{*},j}	& \Sigma_{J^{*},J^{*}}
\end{array}
\right]
\end{eqnarray*}
where $J^{*}$ corresponds to the indices in $\bbX_{J^{*}}$, so $\sigma_{j,j}$ is a scalar, $\Sigma_{j,J^{*}}$ is a $p-1$ dimensional row vector and $\Sigma_{J^{*},J^{*}}$ is a $(p-1)\times (p-1)$ square matrix.
Inverting this partitioned matrix \citep[see, for instance][]{hocking:1996:methods-and-applications-of-linear-models:-regression-and-the-analysis-of-variance} yield:
\begin{align}
	\left[ 
	\begin{array}{cc}
		\sigma_{j,j} & \Sigma_{j,J^{*}} \\
		\Sigma_{J^{*},j} & \Sigma_{J^{*},J^{*}} 
	\end{array}
	\right]^{-1} & =  \left[ 
	\begin{array}{cc}
		\frac{1}{d_{j}^{2}}
		& 
		-\frac{1}{d_{j}^{2}}\beta_{j}
		\\
%		-\frac{1}{d_{j}^{2}}\beta_{j}'
		\bfM_{1}
%		-
%		-(\Sigma_{J^{*}J^{*}}
%		\Sigma_{J^{*}j}\Sigma_{J^{*}J^{*}}^{-1}\Sigma_{J^{*}j})^{-1}
%		\Sigma_{J^{*}J^{*}}^{-1}\Sigma_{J^{*}j} 
		& 
		\bfM_{2}^{-1}
%		\Sigma_{J^{*}J^{*}}
%		-
%		\Sigma_{J^{*}j}\Sigma_{jj}^{-1}\Sigma_{jJ^{*}} 
	\end{array}
	\right],
\label{equation:inverse_of_partitioned_covariance}	
\end{align}
where:
\begin{eqnarray*}
\beta_{j} & = & 
\left(
\begin{array}{ccccccc}
	\beta_{j,1} & \ldots, \beta_{j,j-1}, \beta_{j,j+1}, \ldots, \beta_{j,p}
\end{array}
\right)
=
-\Sigma_{j,J^{*}}\Sigma_{J^{*},J^{*}}^{-1}\in \bbR^{(p-1)},
\\
d_{j} 
& = &  
\sqrt
{\left(\sigma_{j,j}-\Sigma_{j,J^{*}}\Sigma_{J^{*},J^{*}}^{-1}\Sigma_{J^{*},j}\right)}
\in 
\bbR_{+},
\\
\bfM_{2} & = &
		\Sigma_{J^{*},J^{*}}
		-
		\Sigma_{J^{*},j}\sigma_{j,j}^{-1}\Sigma_{j,J^{*}}, 
\\
		\bfM_{1} & = & 
		-
		\bfM_{2}^{-1}\cdot
		\left(\Sigma_{J^{*},J^{*}}^{-1}\Sigma_{J^{*},j}\right)
%		\stackrel{\mbox{symmetry}}{=}
		=
		-\frac{1}{d_{j}^{2}}\beta_{j}',
		\mbox{, (the second equality due to symmetry).}
\end{eqnarray*}
We will focus on the $d_{j}$ and $\beta_{j}$ parameters in what follows.

The parameters $\beta_{j}$ and $d_{j}^{2}$ correspond respectively to the coefficients and the expected value of the squared residuals of the best linear model of $\bbX_{j}$ based on $\bbX_{J^{*}}$, irrespectively of the  distribution of $\bbX$. In what follows, we will let $\beta_{jk}$ denote the coefficient corresponding to $\bbX_{k}$ in the linear model of $\bbX_{j}$ based on $\bbX_{J^{*}}$.
We define:
\begin{itemize}
	\item $\bfD$: a $p\times p$ diagonal matrix with $d_{j}$ along its diagonal and, 
	\item $\bfB$: a $p\times p$ matrix with zeros along its diagonal and off-diagonal terms given by $\beta_{jk}$.
\end{itemize}
Using (\ref{equation:inverse_of_partitioned_covariance}) for $j=1, \ldots, p$ yields:
\begin{eqnarray}
	\Sigma^{-1} & = & \bfD^{-2}\left(\bfI_{p}-\bfB\right)
	\label{equation:precision_matrix_as_function_of_D_B}	
\end{eqnarray}
Since $\Sigma^{-1}$ is symmetric, \eqref{equation:precision_matrix_as_function_of_D_B} implies that the following constraints hold:
\begin{eqnarray}
	\begin{array}{ccccc}
		d_{k}^{2}\beta_{jk} & = & d_{j}^{2}\beta_{kj}, & \mbox{ for } & j,k=1, \ldots, p.
	\end{array}
	\label{equation:symmetry_constraints}
\end{eqnarray}

%In the Gaussian case, the sparsity pattern of $C=\Sigma^{-1}$ determines the conditional dependence structure of the $p$ random variables in $\bbX$ \citep{lauritzen:1996:graphical-models}.
%This fact was explored by  \citet{meinshausen:2006:high-dimensional-graphs-and-variable-selection-with-the-lasso} to estimate the neighborhood of a Gaussian graphical model.
%They estimated the $\beta_{j}$ in each row using the LASSO \citep{tibshirani:1996:lasso}, that is, they obtained:

Equation \eqref{equation:precision_matrix_as_function_of_D_B} shows that the sparsity pattern of $\Sigma^{-1}$ can be inferred from sparsity in the regression coefficients contained in $\bfB$.
\citet{meinshausen:2006:high-dimensional-graphs-and-variable-selection-with-the-lasso} exploit this fact to estimate the neighborhood of a Gaussian graphical model.
They use the LASSO \citep{tibshirani:1996:lasso} to obtain sparse estimates of $\beta_{j}$:
\begin{eqnarray}
\begin{array}{cccl}
\hat{\beta}_{j}(\lambda_{j})
& = & \arg\min\limits_{b_{j}\in\bbR^{p-1}} & \|\bfX_{j} - \bfX_{J^{*}}b_{j}\|^{2} + \lambda_{j}\|b_{j}\|_{1}, \mbox{ for } j = 1, \ldots, p
\label{equation:single_row_lasso_estimates}
\end{array}
\end{eqnarray}
The neighborhood of the node $\bbX_{j}$ was then estimated based on the entries of the $\hat{\beta}_{j}$ that were set to zero.
Minor inconsistencies could occur as the regressions are run separately.
As an example, one could have $\hat{\beta}_{jk}(\lambda_{j})= 0$ and $\hat{\beta}_{kj}(\lambda_{k})\neq0$, which \citet{meinshausen:2006:high-dimensional-graphs-and-variable-selection-with-the-lasso} solve by defining AND and OR rules for defining the estimated neighborhood.

To extend the framework of \citet{meinshausen:2006:high-dimensional-graphs-and-variable-selection-with-the-lasso} to the estimation of precision matrices the parameters $d_{j}^{2}$ must also be estimated and the symmetry constraints in \eqref{equation:symmetry_constraints} must be enforced.
We use a pseudo-likelihood approach \citep{besag:1974:spatial-interaction-and-the-statistical-analysis-of-lattice-systems}
to form a surrogate loss function involving all terms of $\bfB$ and $\bfD$.
For Gaussian $\bbX$, the negative log-likelihood function of $\bbX_{j}$ given $\bbX_{J^{*}}$ is:
\begin{eqnarray}
	\begin{array}{ccccc}
		\log\left[p(\bfX_{j}|\bfX_{J^{*}}, d_{j}^{2}, \beta_{j})\right] 
		& = &
		-\frac{n}{2}\log(2\pi)
		-\frac{n}{2}\log(d_{j}^{2})
		-\frac{1}{2}
		\left(\frac{\|\bfX_{j}-\bfX_{J^{*}}\beta_{j}\|^{2}}{d_{j}^{2}}\right).
	\end{array}
	\label{equation:gaussian_univariate_loglikelihood}
\end{eqnarray}
The parameters $d_{j}^{2}$ and $\beta_{j}$ can be consistently estimated by minimizing (\ref{equation:gaussian_univariate_loglikelihood}). A pseudo-neg-loglikelihood function can be formed as: 
%by multiplying the conditional likelihoods 
\begin{eqnarray}
	\begin{array}{lllll}
		\mcalL(\bfX;\bfD, \bfB) & = & 
		\log\left(\prod_{j=1}^{p}p(\bfX_{j}|\bfX_{J^{*}}, d_{j}^{2}, \beta_{j})\right)
		\\
		& = & 
		-\frac{np}{2}\log(2\pi)		
		-\frac{n}{2}\log\det(\bfD^{2})
		-\frac{1}{2}
		\mbox{tr}\left[\bfX(\bfI_{p}-\bfB')\bfD^{-2}(\bfI_{p}-\bfB)\bfX'\right]
		.
	\end{array}
	\label{equation:gaussian_joint_pseudo_loglikelihood}
\end{eqnarray}
An advantage of the surrogate $\mcalL(\bfX;\bfD, \bfB)$ is that, for fixed $\bfD$, it is a quadratic form in $\bfB$.
To promote sparsity on the precision matrix, we propose using a weighted $\norml{1}$-penalty on $\bfB$:
\begin{align}
	\begin{array}{cccl}
		\left(\hat{\bfD}(\lambda), \hat{\bfB}(\lambda)\right)
		& = &
		\arg\min\limits_{(\bfB,\bfD)} & \{n\log\det(\bfD^2) + 
		\mbox{tr}\left[\bfX(\bfI_{p}-\bfB')\bfD^{-2}(\bfI_{p}-\bfB)\bfX'\right]
		\} + \lambda \|\bfB\|_{w,1}\\
		& & \mbox{s.t.}    &
		 					 \left\{
		                     \begin{array}{llll}
			                 b_{jj} & = & 0 \\
                             d_{kk}^{2}b_{jk} & = & d_{jj}^{2}b_{kj}\\
                             d_{kj}^{2} & = & 0, & \mbox{ for } k\neq j \\
                             d_{jj}^{2} & \ge & 0
                         \end{array}\right.		
	\end{array}
	\label{equation:sparse_pseudo_loglikelihood_estimate}
\end{align}
where $\|\bfB\|_{w, 1} = \sum_{j,k=1}^{p}w_{jk}|b_{jk}|$.
From (\ref{equation:precision_matrix_as_function_of_D_B}), the precision matrix estimate $\hat{\bfC}(\lambda)$ is then given by:
\begin{eqnarray}
	\hat{\bfC}(\lambda) = \hat{\bfD}^{-2}(\lambda)\left[\bfI_{p}-\hat{\bfB}(\lambda)\right].
	\label{equation:splice_precision_matrix_estimate}
\end{eqnarray}

The weights $w_{jk}$ in (\ref{equation:sparse_pseudo_loglikelihood_estimate}) can be adjusted to accommodate differences in scale between the $b_{jk}$ parameters.
In the remainder of this paper, we fix $w_{jk}=1$ for all $j,k$ such that $j\neq k$ (notice that the weights $w_{jj}$ are irrelevant since $b_{jj}=0$ for all $j$).

The main advantage of the pseudo-likelihood estimates as defined in (\ref{equation:sparse_pseudo_loglikelihood_estimate}) is algorithmic.
Fixing $\bfD$, the minimizer with respect to the $\bfB$ parameter is the solution of a $\norml{1}$-penalized least squares problem. 
Hence, for fixed $\bfD$ it is possible to adapt the homotopy/LARS-LASSO algorithm \citep{osborne:2000:lasso_dual, efron:2004:lars} to obtain estimates for all values of $\lambda$.
For each fixed $\bfB$, the minimizer with respect to $\bfD$ has a closed form solution.
The algorithm presented in Section \ref{section:pseudo_likelihood_algorithm} is based on performing alternate optimization steps with respect to $\bfB$ and $\bfD$ to take advantage of these properties.

%At the same time, for a fixed $\bfB$ the minimized $\bfD$ is easily obtained by a closed form solution given by the estimate of the expected value of the squared residuals:
%\begin{eqnarray}
%	\hat{s}_{j} & = & \left[\frac{\left\|\bfX_{j}-\bfX_{J^{*}}b_{j}\right\|^{2}}{n}\right]^{\frac{1}{2}}
%\end{eqnarray}

One drawback of the precision matrix estimate $\hat{\bfC}(\lambda)$ in (\ref{equation:sparse_pseudo_loglikelihood_estimate}) is that it cannot be ensured to be positive semi-definite.
However, from the optimality conditions to \eqref{equation:sparse_pseudo_loglikelihood_estimate} it can be proven that for a large enough $\lambda$, all terms in the $\hat{\bfB}(\lambda)$ matrix are set to zero.
%we point out that, for large enough $\lambda$, $\hat{\bfC}(\lambda)$ is diagonal with positive entries along its diagonal and thus positive semi-definite (eliminating constant $\bfX_{j}$ from the analysis).
For such highly regularized estimates, $\hat{C}(\lambda)$ is diagonal.
Therefore, continuity of the regularization path implies existence of $\lambda^{*}<\inf\{\lambda:\hat{\bfB}(\lambda)=0\}$ for which $\hat{\bfC}(\lambda^{*})$ is diagonal dominant and thus positive semi-definite.

We return to the issue of positive semi-definiteness later in the paper.
We will prove in Section \ref{section:l2_penalization_and_psd} that, when the $\norml{1}$-norm is replaced by the $\norml{2}$-norm, positive semi-definiteness is ensured for any value of the regularization parameter.
That suggests that a penalty similar to the elastic net \citep{zou:2005:elastic-net} can make the estimates along the $\norml{1}$-norm regularization path positive semi-definite for a larger stretch.
In addition, in Section \ref{section:simulation_results_splice} we present evidence that the $\norml{1}$-norm penalized estimates are positive semi-definite for most of the regularization path.

%Unlike the estimates based on the Cholesky representation, however, it is insensitive to the order in which the variables are presented and it imposes sparsity directly on the precision matrix estimate.

%%%%%%%%%%%%%%%%%%%%%%%%%%%%%%%%%%%%%%%%%%%%%%%%%%%%%%%%%%%%%%%%%%%%%%%%%%%%%%%%%%%%%%%%%%%%%%%%%%%%%%%%%%

\subsection{Alternative normalization}
\label{section:pseudo_likelihood_normalized_form}

Before we move on, we present a normalization of the $\bfX$ data matrix that leads to a more natural representation of the precision matrix $\hat{\bfC}$ in terms of $\hat{\bfB}$ and $\hat{\bfD}$ while resulting in more convenient symmetry constraints.

The symmetry constraints in (\ref{equation:symmetry_constraints}) can be rewritten in a more insightful form:
\begin{eqnarray}
	\frac{d_{k}}{d_{j}}\beta_{jk} & = &  \frac{d_{j}}{d_{k}}\beta_{kj}, \mbox{ for all } j\neq k.
	\label{equation:renormalized_symmetry_constraints}
\end{eqnarray}
This alternative representation, suggests that the symmetry constraints can be more easily applied to a renormalized version of the data.
We define:
\begin{eqnarray}
	\label{equation:normalized_parameters_B_D}
	\begin{array}{lllccc}
	  \tilde{\bfX} & = & \bfX \bfD^{-1}, & & & \mbox{ and }\\
	  \tilde{\bfB} & = & \bfD^{-1} \bfB \bfD.
	\end{array}	
\end{eqnarray}
Under this renormalization, $\tilde{\bfB}$ is symmetric, a fact that will be explored in the algorithmic section below to enforce the symmetry constraint within the homotopy algorithm used to trace regularization paths for SPLICE estimates.

Another advantage of this renormalization is that the precision matrix estimate can be written as:
\begin{eqnarray}
	\begin{array}{lllccc}
	  \hat{\bfC}(\lambda) & = & \hat{\bfD}^{-1}\left[\bfI_{p}-\hat{\tilde{\bfB}}(\lambda)\right]\hat{\bfD}^{-1},
	\end{array}	
\end{eqnarray}
making 
%it evident that $\bfD$ does not play a part in determining 
the analysis of the positive semi-definiteness of $\hat{\bfC}(\lambda)$ easier: $\hat{\bfC}(\lambda)$ is positive semi-definite if and only if $\bfI_{p}-\hat{\tilde{\bfB}}(\lambda)$ is positive semi-definite.
This condition is satisfied as long as the eigenvalues of $\hat{\tilde{\bfB}}(\lambda)$ are smaller than $1$.

%%%%%%%%%%%%%%%%%%%%%%%%%%%%%%%%%%%%%%%%%%%%%%%%%%%%%%%%%%%%%%%%%%%%%%%%%%%%%%%%%%%%%%%%%%%%%%%%%%%%%%%%%%

\subsection{Comparison of exact and pseudo-likelihoods in the Gaussian case}
\label{section:comparison_exact_pseudo_likelihood}

\citet{banerjee:2005:sparse-covariance-selection-via-robust-maximum-likelihood-estimation}, 
\citet{yuan:2007:model-selection-and-estimation-in-the-gaussian-graphical-model}, \citet{banerjee:2007:model-selection-through-sparse-maximum-likelihood-estimation-for-multivariate-gaussian-or-binary} and \citet{friedman:2008:sparse-inverse-covariance-estimation-with-the-graphical-lasso}
have considered estimates defined as the minimizers of the exact likelihood penalized by the $\norml{1}$-norm of the off-diagonal terms of the precision matrix $\bfC$:
\begin{eqnarray}
	\begin{array}{ccll}
		\hat{\bfC}_{exact}(\lambda)
		& = &
		\arg\min_{\bfC} & n\log\det(\bfC) + \mbox{tr}\left[\bfX \bfC\bfX'\right] + \lambda \|\bfC\|_{1}\\
		& & \mbox{s.t.}     & \bfC \mbox { is symmetric positive semi-definite.}
	\end{array}
	\label{equation:sparse_exact_loglikelihood_estimate}
\end{eqnarray}

A comparison of the exact and pseudo-likelihood functions in terms of the $(\bfD, \tilde{\bfB})$ parametrization suggests when the approximation will be appropriate.
In terms of $(\bfD, \tilde{\bfB})$, the pseudo-likelihood function in \eqref{equation:gaussian_joint_pseudo_loglikelihood} is:
\begin{eqnarray}
	\begin{array}{lllll}
		\mcalL(\bfX;\bfD, \bfB) & = & 
		-\frac{np}{2}\log(2\pi)		
		-\frac{n}{2}\log\det(\bfD^{2})
		-\frac{1}{2}
		\mbox{tr}\left[\tilde{\bfX}(\bfI_{p}-\tilde{\bfB})^{2}\tilde{\bfX}'\right]
		.
	\end{array}
	\label{equation:renormalized_joint_pseudo_loglikelihood}
\end{eqnarray}
In the same parametrization, the exact likelihood function is:
\begin{align}
	\begin{array}{rcl}
		\mcalL_{exact}(\bfX;\bfD, \bfB)
		& = &
		-\frac{np}{2}\log(2\pi)
		-\frac{n}{2}\log\det\left[\left(\bfI_{p}-\tilde{\bfB}\right)\right]
		\\
		& & {} 		
		-\frac{n}{2}\log\det(\bfD^{2})
		-\frac{1}{2}
		\mbox{tr}\left[\tilde{\bfX}\left(\bfI_{p}-\tilde{\bfB}\right)\tilde{\bfX}'\right]
	\end{array}
	\label{equation:gaussian_joint_exact_loglikelihood}
\end{align}
A comparison between (\ref{equation:renormalized_joint_pseudo_loglikelihood}) and (\ref{equation:gaussian_joint_exact_loglikelihood}) reveals two differences in the exact and pseudo neg-loglikelihood functions.
First, the exact expression involves one additional $\log\det\left[\left(\bfI_{p}-\tilde{\bfB}\right)\right]$ term not appearing in the surrogate expression.
Secondly, in the squared deviation term, the surrogate expression has the weighting matrix  $\left[\left(\bfI_{p}-\tilde{\bfB}\right)\right]$ squared in the comparison to the exact likelihood.
For $\tilde{\bfB}=\bfzero$, the $\log\det$ vanishes and the weighting term equals identity and thus idempotent.
Since these two functions are continuous in $\tilde{\bfB}$, we can expect this approximation to work better the smaller the off-diagonal terms in $\tilde{\bfB}$.
In particular, in the completely sparse case, the two functions coincide and the approximation is exact.
%We follow up this discussion on the quality of the pseudo-likelihood approximation in our experimental section below (Section \ref{section:simulation_results_splice}).

%%%%%%%%%%%%%%%%%%%%%%%%%%%%%%%%%%%%%%%%%%%%%%%%%%%%%%%%%%%%%%%%%%%%%%%%%%%%%%%%%%%%%%%%%%%%%%%%%%%%%%%%%%

\subsection{Properties of the pseudo-likelihood estimate}
\label{section:pseudolikelihood_estimate_properties}

In the classical setting where $p$ is fixed and $n$ grows to infinity, the unregularized pseudo-likelihood estimate $\hat{\bfC}(0)$ is clearly consistent. 
The unconstrained estimates for $\beta_{j}$ and $d_{j}^{2}$ are all consistent. 
In adition, the symmetry constraints we impose are observed in the population version and thus introduce no asymptotic bias in the (symmetry) constrained estimates.
That in conjunction with the results from \citet{knight:2000:asymptotics} can be used to prove that, as long as $\lambda = o_{p}(n)$ the $\norml{1}$-penalized estimates in (\ref{equation:sparse_pseudo_loglikelihood_estimate}) are consistent.

Still in the classical setting, \citet{yuan:2007:model-selection-and-estimation-in-the-gaussian-graphical-model} point out that the unpenalized estimate $\hat{\bfC}(0)$ is not efficient as it does not coincide with the maximum likelihood estimate.
However, the comparison of the exact and pseudo-likelihoods presented in Section \ref{section:comparison_exact_pseudo_likelihood} suggests that the loss in efficiency should be smaller the sparser the true precision matrix.
In addition, the penalized pseudo-likelihood estimate lends itself better for path following algorithms and can be used to select an appropriate $\lambda$ while simultaneously providing a good starting point for algorithms computing the exact solution.

One interesting question we will not pursue in this paper concerns the quality of the pseudo-likelihood approximation in the non-classical setting where $p_{n}$ is allowed to grow with $n$.
In that case, the classical efficiency argument favoring the exact maximum likelihood over the pseudo-likelihood approximation no longer holds.
To the best of our knowledge, it is an open question whether the penalized exact ML has advantages over a pseudo-likelihood approximation in the non-parametric case ($p_{n}$ growing with $n$).

%%%%%%%%%%%%%%%%%%%%%%%%%%%%%%%%%%%%%%%%%%%%%%%%%%%%%%%%%%%%%%%%%%%%%%%%%%%%%%%%%%%%%%%%%%%%%%

\subsection{Penalization by $\norml{2}$-norm and Positive Semi-definiteness}
\label{section:l2_penalization_and_psd}

As mentioned above,the algorithm we propose in Section \ref{section:pseudo_likelihood_algorithm} suffers from the drawback of not enforcing a positive semi-definite constraint.
Imposing such constraint is costly in computational terms and would slow down our path following algorithm.
As we will see in the experimental Section \ref{section:simulation_results_splice} below, the unconstrained estimate is positive semi-definite for the greater part of the regularization path even in nearly singular designs.

Before we review such experimental evidence, however, we study an alternative penalization that does result in positive semi-definite estimates for all levels of regularization.
Consider the penalized pseudo-likelihood estimate defined by:
\begin{eqnarray}
		\begin{array}{cccl}
			\hat{\tilde{\bfB}}_{2}(\lambda_{2})
			& = &
			\arg\min\limits_{\tilde{\bfB}} & \left\{
			\mbox{tr}\left[(\bfI_{p}-\tilde{\bfB})\tilde{\bfX}'\tilde{\bfX}(\bfI_{p}-\tilde{\bfB}')\right]
			+ \lambda_{2} \cdot\mbox{tr}\left[\tilde{\bfB}'\tilde{\bfB}\right]\right\}\\
			& & \mbox{s.t.}    &
			 					 \left\{
			                     \begin{array}{llll}
				                 b_{jj} & = & 0 \\
	                             \tilde{b}_{jk} & = & \tilde{b}_{kj}\\
	                         \end{array}\right.		
		\end{array}
		\label{equation:ridged_pseudo_loglikelihood_estimate}
\end{eqnarray}

Our next result establishes that the estimate defined by  is positive semi-definite along its entire path:
\begin{theorem}
	Let $\hat{C}(\lambda_{2})=\hat{\bfD}_{2}(\lambda_{2})^{-1}\left(\bfI_{p}-\hat{\tilde{\bfB}}_{2}(\lambda_{2})\right)\hat{\bfD}_{2}(\lambda_{2})^{-1}$ be the precision matrix estimate resulting from (\ref{equation:ridged_pseudo_loglikelihood_estimate}).
	For any $\hat{\bfD}_{2}(\lambda_{2})$ and $\lambda_{2}>0$, the precision matrix estimate $\hat{\bfC}_{2}(\lambda_{2})$ is positive semi-definite.
\label{theorem:ridged_pseudolikelihood_estimate_is_psd}
\end{theorem}
%Interestingly, a by-product of our proof is that the shrunken covariance estimate defined in \citet{ledoit:2004:large_covariance_matrix} is the inverse of the precision matrix estimate at a particular level of regularization.
%This result is detailed in the Appendix.

This result suggests that the $\norml{2}$-penalty may be useful in inducing positive semi-definiteness.
%In that sense, an estimate that is penalized by both the $\norml{2}$- and $\norml{1}$-penalties can have improved performance in what refers to positive semi-definiteness:
So an estimate incorporating both $\norml{2}$- and $\norml{1}$-penalties, 
\begin{align}
\begin{array}{cccl}
\hat{\tilde{\bfB}}_{EN}(\lambda_{2}, \lambda_{1})
& = &
\arg\min\limits_{\tilde{\bfB}} & \left\{
\mbox{tr}\left[(\bfI_{p}-\tilde{\bfB})\tilde{\bfX}'\tilde{\bfX}(\bfI_{p}-\tilde{\bfB}')\right]
+ \lambda_{2} \cdot\mbox{tr}\left[\tilde{\bfB}'\tilde{\bfB}\right]
+ \lambda_{1} \cdot\left|\tilde{\bfB}'\tilde{\bfB}\right|_{w, 1}
\right\}\\
& & \mbox{s.t.}    &
\left\{
\begin{array}{llll}
b_{jj} & = & 0 \\
\tilde{b}_{jk} & = & \tilde{b}_{kj}\\
\end{array}\right.,
\end{array}
\label{equation:elastic_net_pseudo_loglikelihood_estimate}
\end{align}
can have improved performance, over the $\norml{1}$-penalty alone, in terms of positive semi-definiteness.
The subscript $EN$ is a reference to the Elastic Net penalty of \citet{zou:2005:elastic-net}.
In our experimental section below (Section \ref{section:simulation_results_splice}), we have came across non-positive semi-definite precision matrix estimate in none other than a few replications.
We hence left a detailed study of the Elastic Net precision matrix estimate defined in \eqref{equation:elastic_net_pseudo_loglikelihood_estimate} for future research.

%[TODO: Can we prove that for any $\lambda_{1}$, there exists $\lambda_{2}$ large enough such that estimate is PSD?]

%%%%%%%%%%%%%%%%%%%%%%%%%%%%%%%%%%%%%%%%%%%%%%%%%%%%%%%%%%%%%%%%%%%%%%%%%%%%%%%%%%%%%%%%%%%%%%

\section{Algorithms}
\label{section:pseudo_likelihood_algorithm}

We now detail an iterative algorithm for computing the regularization path for SPLICE estimates.
The computational advantage of that approximate loss function used in \eqref{equation:sparse_pseudo_loglikelihood_estimate} follows from two facts:
\begin{itemize}
\item For a fixed $\bfD$, the optimization problem reduces to that of tracing a constrained LASSO path;
\item For a fixed $\bfB$, the optimizer $\hat{\bfD}$ has a closed form solution;
\end{itemize}

Based on these two facts, we propose an iterative algorithm for obtaining estimates $\hat{\bfD}(\hat{\lambda})$
and $\hat{\bfB}(\hat{\lambda})$.
Within each step, the path for $\hat{\bfD}$ and $\hat{\bfB}$ as a function of the regularization parameter $\lambda$ is traced and the choice of the regularization parameter $\lambda$ is updated.
We now describe the steps of our SPLICE algorithm.
\begin{enumerate}
\item Let $k=1$ and $\hat{\bfD}_{0} = \diag\left(\frac{\|\bfX_{j}\|_{2}^{2}}{n}\right)_{j=1}^{p}$.
\item Repeat until \textit{convergence}:
\begin{enumerate}

\item Set $\tilde{\bfX} = \bfX \hat{\bfD}_{k-1}^{-1}$ and use a path-tracing algorithm for the regularized problem:
\begin{eqnarray*}
	\begin{array}{cccl}
	\hat{\tilde{\bfB}}(\lambda)
	& = &
	\arg\min\limits_{\bfB}
	&
	\left\{
	\trace\left[\tilde{\bfX}(\bfI-\tilde{\bfB}')(\bfI-\tilde{\bfB})\tilde{\bfX}'\right]+\lambda \cdot \|\tilde{\bfB}\|_{\tilde{w},1}\right\}
	\\
	& & \mbox{s.t.} & \tilde{b}_{jk}=\tilde{b}_{kj} \\
	& &             & \tilde{b}_{jj}=0
	\end{array}
\end{eqnarray*}

\item Compute the corresponding path for $\bfD$:
\begin{eqnarray*}
	\hat{\bfD}_{k}(\lambda) = \diag\left(\frac{\|\left(\bfI_{p}-\hat{\bfB}(\lambda)\right)\bfX_{j}\|_{2}^{2}}{n}\right)_{j=1}^{p}
\end{eqnarray*}

\item Select a value for the regularization parameter $\hat{\lambda}_{k}$;

\item Set $\hat{\bfB}_{k} = \hat{\bfB}(\hat{\lambda}_{k})$ and $\hat{\bfD}_{k} = \hat{\bfD}(\hat{\lambda}_{k})$;

\end{enumerate}

\item Return the estimate 
$\hat{\bfC} = \hat{\bfD}_{k}^{-1}(\bfI-\hat{\tilde{\bfB}}_{k})\hat{\bfD}_{k}^{-1}$;

\end{enumerate}

%Notice that the problem in step 2a refers to a path in a normalized version of the problem. 
%The vector of weights $w$ in our weighted version of the $\norml{1}$-penalty allows us to adjust the penalty so that $\|\tilde{\bfB}\|_{w, 1} = \|\bfB\|_{1}$.
%Alternatively, $w$ can be used to allow some degree of adaptiveness in the penalization.
%In this paper, we stick to the penalty $\|\bfB\|_{1}$.

A subtle but important detail is the weighting vector $\tilde{w}$ used when applying the penalty in step 2.(a).
Since the path is traced in terms of the normalized $\tilde{\bfB}$ parameter instead of $\bfB$, a correction must be made in these weights.
This can be determined by noticing that:
\begin{eqnarray*}
	\|\bfB\|_{w, 1}
	=
	\sum_{j,k=1}^{p}w_{jk} \left|b_{jk}\right|
	=
	\sum_{j,k=1}^{p}w_{jk}\frac{{d}_{j}}{{d}_{k}} \left|\frac{{d}_{k}}{{d}_{j}}b_{jk}\right|	
	= 
	\sum_{j,k=1}^{p}\tilde{w}_{jk} \left|\tilde{b}_{jk}\right|
	=
	\|\tilde{\bfB}\|_{\tilde{w}, 1},
\end{eqnarray*}
as long as $\tilde{w}_{jk}=w_{jk}\frac{{d}_{j}}{{d}_{k}}$.
As mentioned before, we fix $w_{jk}=1$ throughout this paper, so we set $\tilde{w}_{jk}=w_{jk}\frac{{d}_{j}}{{d}_{k}}$ in our experiments. 
Of course, ${d}_{j}$ are unknown so the current estimate is plugged-in every time step 2.(a) is executed.

%Adaptive penalties in the same spirit of the adaptive LASSO \citep{zou:2006:adaptive_lasso} can be proposed by setting different updating policies for the weights.
%For instance, in cases with very disparate variances across the $p$ random variables in $\bfX$, an argument can be made for setting ${w}_{jk}=\frac{d_{k}}{d_{j}}$ so that $\tilde{w}_{jk}=1$.
%We leave the investigation of adaptive weighting as a theme for future research.
In the remainder of this section, we show how to adapt the homotopy/LARS-LASSO algorithm to enforce the symmetry constraints $d_{k}^{2}b_{jk}=d_{j}^{2}b_{kj}$ along the regularization path, how to select $\hat{\bfB}$ and $\hat{\bfD}$ from the path, and discuss some convergence issues related to the algorithm above. 

\subsection{Enforcing the symmetry constraints along the path}
\label{section:algorithm_symmetry_constraint}

%As explained in Section \ref{section:pseudolikelihood_loss_function} above, the $\bfB$ matrix must satisfy the constraint $d_{j}^{2}\beta_{kj}=d_{k}^{2}\beta_{jk}$.
%Given a fixed $\hat{\bfD}_{k-1}$, the regularization path for $\hat{\bfB}_{k}$ is defined by:
The expression defining the regularization path in step 2.(a) of the SPLICE algorithm above can be rewritten as:
\begin{eqnarray}
	\begin{array}{ccll}
		\hat{\bfB}(\lambda)
		& = &
		\arg\min\limits_{\bfB} & 
		\mbox{vec}\left(\tilde{\bfX}(\bfI_{p}-\bfB')\right)'
		\mbox{vec}\left(\tilde{\bfX}(\bfI_{p}-\bfB')\right)
		+ \lambda\|\bfB\|_{w, 1}
		\\
		& & \mbox{s.t.}    & b_{jj} = 0 \\
		& &                & b_{jk} = b_{kj},		
	\label{equation:quadratic_l1_penalized_regularization_path}
	\end{array}
\end{eqnarray}
which is a quadratic form in $\bfB$ penalized by a weighted version of its $\norml{1}$-norm.
To enforce the equality restriction in (\ref{equation:quadratic_l1_penalized_regularization_path}), we massage the data into an appropriate form so the homotopy/LARS-LASSO algorithm can be used in its original form.

The optimization problem in \eqref{equation:quadratic_l1_penalized_regularization_path} corresponds to a constrained version of a penalized regression of modified response ($\bfY$) and predictors ($\bfZ$) given by:
\begin{eqnarray*}
\begin{array}{cclcccl}
    \bfY 
	& = &
%	\left[
%	\hat{\bfD}_{k-1}^{-1}\otimes \bfI_{n}
%	\right]	
	\left[
	\begin{array}{c}
		\tilde{\bfX}_{1} \\
		\tilde{\bfX}_{2} \\
		\vdots \\
%		\tilde{\bfX}_{p-1}		\\
		\tilde{\bfX}_{p}		
	\end{array}
	\right]%_{np\times 1}
	&
	\mbox{ and }
	\\
    \bfZ
	& = &
%	\left[
%	\hat{\bfD}_{k-1}^{-1}\otimes \bfI_{n}
%	\right]	
	\left[
	\begin{array}{cccccccccc}
		\tilde{\bfX}_{1^{*}} & \bfzero & \bfzero & \cdots & \bfzero \\
		\bfzero & \tilde{\bfX}_{2^{*}} & \bfzero & \cdots & \bfzero \\
		\bfzero & \bfzero & \tilde{\bfX}_{3^{*}} & \cdots & \bfzero \\
		\vdots  & \vdots  & \vdots  & \ddots & \vdots \\
		\bfzero & \bfzero & \bfzero & \cdots & \tilde{\bfX}_{p^{*}}
    \end{array}
	\right]%_{np\times \frac{p(p-1)}{2}}	 
\end{array}
\end{eqnarray*}

Since the ``model'' for $\bfY$ given $\bfZ$ is additive, we can force $b_{jk}=b_{jk}$ by creating a modified design matrix $\tilde{\bfZ}$ where the columns corresponding to $b_{jk}$ and $b_{kj}$ are summed into a single column.
More precisely, the column corresponding to $\tilde{b}_{jk}$ with $j<k$ in the $\tilde{\bfZ}$ design matrix has all elements set to zero except for the rows corresponding to $\bfX_{k}$ and  ${\bfX}_{j}$ in the $\bfY$ vector.
These rows must be set to $\tilde{\bfX}_{j}$ and $\tilde{\bfX}_{k}$ respectively:
\begin{eqnarray*}
\begin{array}{cclcccl}
    \tilde{\bfZ}
	& = &
%	\left[
%	\hat{\bfD}_{k-1}^{-1}\otimes \bfI_{n}
%	\right]	
	\left[
	\begin{array}{cccccccccc}
		\tilde{\bfX}_{2} & \tilde{\bfX}_{3} & \cdots & \tilde{\bfX}_{p-1} & \tilde{\bfX}_{p} & \bfzero & \cdots & \bfzero & \cdots
		& \bfzero\\
		\tilde{\bfX}_{1} & \bfzero          & \cdots & \bfzero          &\bfzero          & \tilde{\bfX}_{3} & \cdots & \tilde{\bfX}_{p} & \cdots & \bfzero\\
		\bfzero & \tilde{\bfX}_{1}          & \cdots & \bfzero          &\bfzero          & \tilde{\bfX}_{2} & \cdots & \bfzero & \cdots & \bfzero\\
		\vdots \\
		\bfzero & \bfzero & \cdots & \tilde{\bfX}_{1} & \bfzero & \cdots & \bfzero          &\bfzero & \cdots & \tilde{\bfX}_{p} \\
		\bfzero & \bfzero & \cdots & \bfzero & \tilde{\bfX}_{1} & \cdots & \bfzero          &\bfzero & \cdots & \tilde{\bfX}_{p-1} 				
	\end{array}
	\right]%_{np\times \frac{p(p-1)}{2}}	 
\end{array}
\end{eqnarray*}

The path for the constrained $\tilde{\bfB}(\lambda)$ can then be traced by simply applying a weighted version of the LASSO algorithm to $\bfY$ and $\tilde{\bfZ}$.
We emphasize that, even though the design matrix $\tilde{\bfZ}$ has $O(np^{3})$ elements, it is extremely sparse. It can be proved that it contains only $O(np^{2})$ non-zero terms.
It is thus crucial that the implementation of the homotopy/LARS-LASSO algorithm used to trace this path make use of the sparsity of the design matrix.

\subsection{Computational complexity of one step of the algorithm}

The algorithm proposed above to trace the regularization path of $\hat{\tilde{\bfB}}(\lambda)$ uses the homotopy/LARS-LASSO algorithm with a modified regression with $\frac{p(p-1)}{2}$ regressors and $np$ observations.
The computational complexity of the $k$-th step of the homotopy algorithm in step 2.(a) is of order $O(a_{k}^{2} + a_{k}np)$, where $a_{k}$ denotes the number of non-zero terms on the upper triangular, off-diagonal of $\tilde{\bfB}(\lambda_{k})$ \citep[a detailed analysis is presented in ][]{efron:2004:lars}.
Clearly, the computational cost increases rapidly as more off-diagonal terms are added to the precision matrix.

When $p$ grows with $n$ keeping a constant ratio $p/n$, we find it plausible that most selection criteria will pick estimates at the most regularized regions of the path: a data sample can only support so many degrees of freedom.
Thus, incorporating early stopping criteria into the path tracing at step 2.(a) of the SPLICE algorithm can greatly reduce the computational cost of obtaining the path even further without degrading the statistical performance of the resulting estimates.

%It follows from the complexity analysis presented in \citet{efron:2004:lars} that the complexity of the $k$-th step of the homotopy algorithm is of order $O(a_{k}^{2} + a_{k}np)$, where $a_{k}$ denotes the number of non-zero terms on the upper triangular, off-diagonal of $\tilde{\bfB}(\lambda_{k})$ at the $k$-th step of the homotopy algorithm.

If one insists in having the entire precision matrix path, the SPLICE algorithm is still polynomial in $p$ and $n$. 
In the case where no variables are dropped and the variables are added one at a time, the complexity of the first $K$ steps of the path is given by $O(K^{3} + K^{2}np)$.
Under the same assumptions and setting $K=0.5\cdot p(p-1)$, the SPLICE algorithm has cost of order $O(p^{6}+np^{5})$ to compute the entire path of solutions to the pseudo-likelihood problem.
As a comparison, an algorithm presented by \citet{banerjee:2005:sparse-covariance-selection-via-robust-maximum-likelihood-estimation} has a cost of  order $O(\frac{p^{4.5}}{\eta})$ to compute an approximate solution for the penalized exact likelihood at a single point of the path, where $\eta$ represents the desired level of approximation.

\subsection{Selection criteria for $\lambda$:}
\label{section:lambda_selection_criteria}

Different criteria can be used to pick a pair $\bfB, \bfD$ from the regularization path (cf. 2(c) in the SPLICE algorithm).
In the experiment section below, we show the results of using Akaike's Information Criterion \citep[AIC, ][]{akaike:1973:information-criterion}, the Bayesian information Criterion \citep[BIC, ][]{schwartz:1978:bic} and a corrected version of the AIC criterion \citep[$\aicc$,][]{hurvich:1998:smoothing-parameter-selection-in-nonparametric-regression-using-an-improved-akaike-information-criterion} using the unbiased estimate of the degrees of freedom of the LASSO along the regularization path.
More precisely, we set:
\begin{eqnarray*}
\begin{array}{ccc}
\hat{\lambda} & = & 
\arg\min_{\lambda}\mathcal{L}_{exact}(\bfX, \hat{\bfD}, \hat{\bfB}(\lambda))+ K(n, \widehat{\mbox{df}}(\hat{\bfC}(\lambda))
\end{array}
\end{eqnarray*}
where:
\begin{eqnarray*}
K(n, \widehat{\mbox{df}}(\hat{\bfC}(\lambda))) & = & 
\left\{
\begin{array}{lll}
2\widehat{\mbox{df}}(\hat{\bfC}(\lambda)), & \mbox{ for AIC} , \\	
\frac{1+\frac{\widehat{\mbox{df}}(\hat{\bfC}(\lambda))}{n}}{1-\frac{\widehat{\mbox{df}}(\hat{\bfC}(\lambda))+2}{n}}, & \mbox{ for $\aicc$} , \\	
\log(n)\widehat{\mbox{df}}(\hat{\bfC}(\lambda)), & \mbox{ for BIC} , \\	
\end{array}
\right.
\end{eqnarray*}
As our estimate of the degrees of freedom along the path, 
we follow \citet{zou:2004:lasso_dfs} and
use $p$ plus the number of non-zero terms in the upper triangular section of $\hat{\bfB}(\lambda)$, that is,
\begin{eqnarray}
	\label{equation:dfs}
	\widehat{\mbox{df}}(\lambda) & = & p + \left|\left\{(i, j): i<j \mbox{ and } \hat{\bfB}_{ij}(\lambda)\neq 0\right\}\right|.
\end{eqnarray}
The $p$ term in expression \eqref{equation:dfs} stems from the degrees of freedom used for estimating the means.

\subsection{Stopping criterion}
\label{section:pseudolikelihood_algorithm_convergence}

%Fixing $\lambda$, the empirical risk evaluated at $\hat{D}_{k+1}(\lambda), \hat{B}_{k+1}(\lambda)$ is no larger than the empirical risk for $\hat{D}_{k}(\lambda), \hat{B}_{k}(\lambda)$ for any $k$.
%Since the pseudo-likelihood empirical risk is convex in $(\bfB, \bfD)$, the convergence of SPLICE algorithm is ensured.

We now determine the convergence criterion we used to finish the loop started in step (2) of the SPLICE algorithm.
To do that, we look at the variation in the terms of the diagonal matrix $D$.
We stop the algorithm once:
\begin{eqnarray*}
	\max_{1\le j \le p}\left\{\log\left(\frac{[\hat\bfD_{k+1}]_{j}}{[\hat\bfD_{k}]_{j}}\right)\right\} 
	& < &  
	10^{-2},
\end{eqnarray*}
or when the number of iterations exceeds a fixed number $Q$, whichever occurs first.

We have also observed that letting $\lambda_{k+1}$ differ from $\lambda_{k}$ often resulted in oscillating estimates caused solely by small variations in $\lambda$ from one step to the next. 
To avoid this issue, we fixed the regularization parameter $\lambda$ after a number $M<Q$ of ``warm-up'' rounds.

We have set $M=6$ and $Q=100$ for the simulations in Section \ref{section:simulation_results_splice}.
For most cases, the selected value of $\lambda$ and the $\hat{\bfD}$ matrix had become stable and the algorithm had converged before either the maximum number $M$ of ``warm-up'' steps was reached.

%%%%%%%%%%%%%%%%%%%%%%%%%%%%%%%%%%%%%%%%%%%%%%%%%%%%%%%%%%%%%%%%%%%%%%%%%%%%%%%%%%%%%%%%%%%%%%

\section{Numerical Results}
\label{section:simulation_results_splice}

In this section, we compare the results obtained by SPLICE in terms of estimation accuracy and model selection performance to other covariance selection methods, namely the Cholesky covariance selection procedure proposed by \citet{huang:2006:covariance-matrix-selection-and-estimation-via-penalized-normal-likelihood} (Cholesky) and the $\ell_{1}$-penalized maximum likelihood estimate studied previously by \citet{yuan:2007:model-selection-and-estimation-in-the-gaussian-graphical-model, banerjee:2005:sparse-covariance-selection-via-robust-maximum-likelihood-estimation} and \citet{friedman:2008:sparse-inverse-covariance-estimation-with-the-graphical-lasso} (Exact PML).
We compare the effectiveness of three different selection criteria --  AIC, $\aicc$ and BIC (see Section \ref{section:lambda_selection_criteria})-- in picking an estimate from the SPLICE, Cholesky and Penalized Exact Log-Likelihood regularization paths.
We also compare the model selection performance over the entire regularization path using ROC curves (for details, see \ref{section:covariance_selection_comparison}).
Finally, we study the positive semi-definiteness of the SPLICE estimates along the regularization path in a near-singular design.

\subsection{Comparison of SPLICE to alternative methods}
\label{section:simulation_results_comparison}

Figure \ref{figure:simulated_topologies} shows the graphical models corresponding to the simulated data sets we will be using to compare SPLICE, Cholesky and Exact PML in terms of estimation accuracy and covariance selection.
All designs involve a 15-dimensional zero-mean Gaussian random vector ($p=15$) with precision matrix implied by the graphical models shown in Figure \ref{figure:simulated_topologies}.
A relatively small sample size is used to emulate the effect of high-dimensionality.
For all comparisons, the estimates are computed based on either 20 or 1,000 independent samples from each distribution (small sample case: $n=20$, large sample case: $n=1,000$).
The results presented here are based on $r=200$ replications for each case.

%\afterpage{\clearpage\input{tables/ar_cov_froeb_table}}
\renewcommand{\figurewidth}{1.00}
\renewcommand{\raiselegendtext}{0.5cm}
\begin{figure}[p]
\begin{center}
\includegraphics[width=\figurewidth\textwidth]{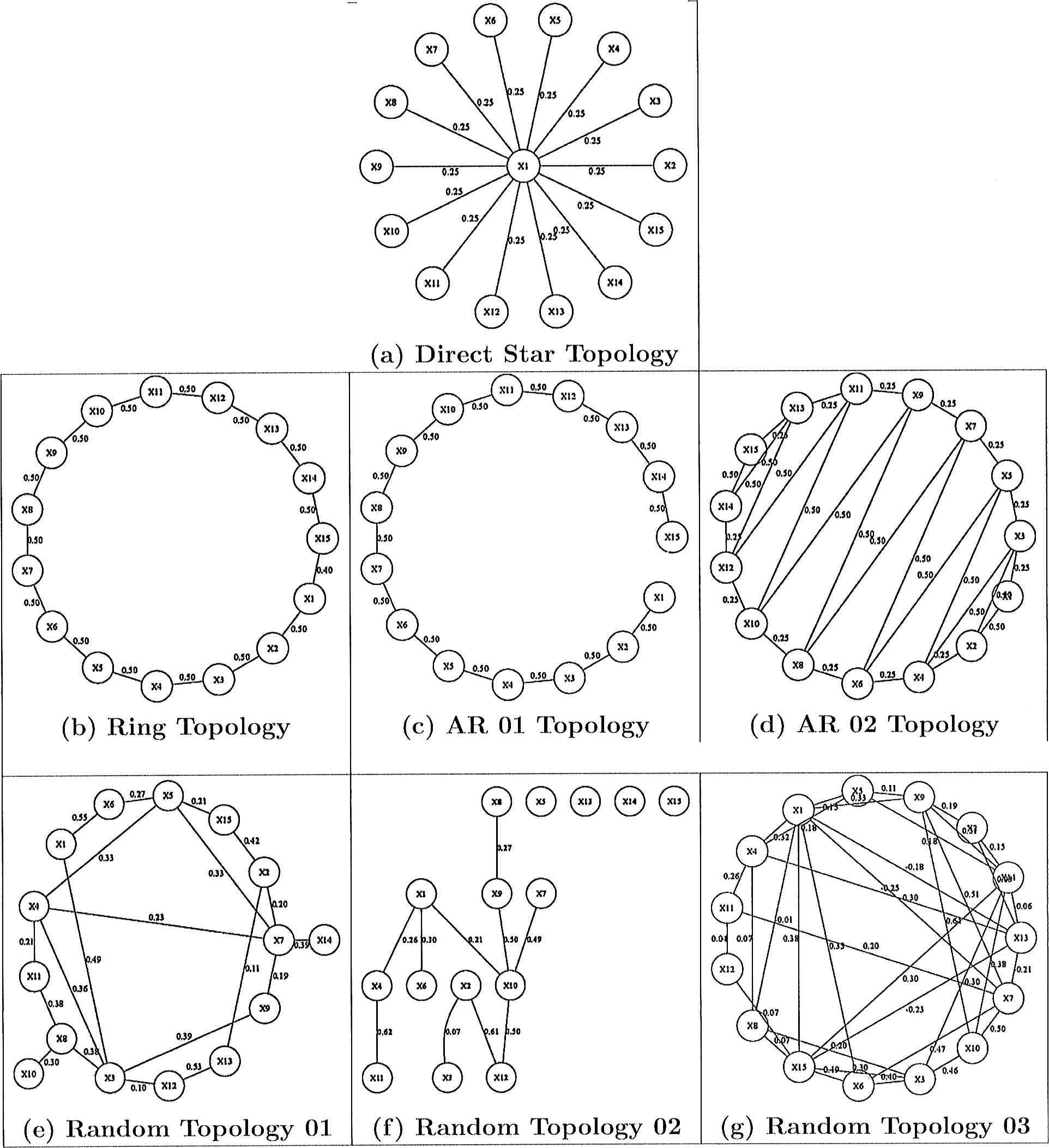}
\caption
{
\label{figure:simulated_topologies}
{\bfseries Simulated cases:}
In this table, the topology of the precision matrix for all simulated cases is shown.
In each case, the precision matrix is normalized to have unit diagonal.
The edges show the value of $c_{ij}$ whenever it differs from zero.
}
\end{center}
\end{figure}

Before we show the results, a few words on our choice of simulation cases:
\begin{itemize}
  \item The star model (see Figure \ref{figure:simulated_topologies} panel a) provides an interesting example where the ordering of the variables can have a great influence in the results of an order-dependent method such as Cholesky. In the \textit{Direct Star} case, the hub variable is the first entry in the $15$-dimensional random vector (as shown in the figure). Conditioning on the first variable ($X_{1}$) makes all other variables independent ($X_{2}, \ldots, X_{15}$). Meanwhile, in the \textit{inverted star} topology, the hub variable is put at the last position of the $15$-dimensional random vector. As a result, no conditional independence is present until the last variable is added to the conditioning set.

  \item In the ``AR-like'' family of models (see Figure \ref{figure:simulated_topologies} panels b, c and d) each $15$-dimensional random vector corresponds (panels c and d) or is very similar (panel b) to 15 observations from an auto-regressive process. This family of models tends to give some advantage to the Cholesky procedure as the ordering of the variables within the vector contain some information about the dependency structure among the variables. The cases in this family are loosely based on some of the simulation designs used in \citet{yuan:2007:model-selection-and-estimation-in-the-gaussian-graphical-model};

  \item The ``random'' designs (see Figure \ref{figure:simulated_topologies} panels e, f and g) were obtained by randomly choosing precision matrices as described in Appendix \ref{section:sampling_precision_matrices}. We used these designs to make sure our results are valid in somewhat less structured environments.

\end{itemize}

\subsubsection{Estimation accuracy of SPLICE, Cholesky and Exact PML}

  We evaluate the accuracy of the precision matrix estimates according to following four metrics.
  \begin{itemize}
    \item The quadratic loss of the estimated precision matrix, defined as
    \begin{eqnarray*}
      \Delta_{2}(\bfC, \hat{\bfC})
      & = & \trace\left(\bfC\hat{\bfC}^{-1}-\bfI_{p}\right)^{2}.
    \end{eqnarray*}	
    
    \item The entropy loss at the estimated precision matrix, defined as
      \begin{eqnarray*}
        \Delta_{e}(\bfC, \hat{\bfC})
        & = & 
        \trace \left(\bfC\hat{\bfC}^{-1}\right)-\log\left(\bfC\hat{\bfC}^{-1}\right)-n.
      \end{eqnarray*}
      
    \item The spectral norm of the deviation of the estimated precision matrix $\left(\hat{\bfC}-\bfC\right)$  where the spectral norm of a square matrix $\bfA$ is defined as $\|\bfA\|_{2}  =  \sup_{x}\frac{\|x'\bfA x\|^{2}}{\|x\|^{2}}$.

    \item The spectral norm of the deviation of the estimated covariance matrix $\left(\hat{\Sigma}-\Sigma\right)$.    
\end{itemize}

  For each of Cholesky, SPLICE and Exact PML, we compute estimates taken from their paths using the selection criteria mentioned in \ref{section:lambda_selection_criteria}: AIC, $\aicc$ and BIC.
The Cholesky and SPLICE estimates are chosen from the breakpoints of their respective regularization paths.
The path tracing algorithm we have used for the Cholesky estimate is sketched in Appendix \ref{appendix:cholesky_path_tracing}.
The path following algorithm for SPLICE is the one described in Section \ref{section:pseudo_likelihood_algorithm}.
The Exact {ML estimate is chosen by minimizing the selection criterion over an equally spaced $500$-point $\lambda$-grid between $0$ and the maximum absolute value of the off-diagonal terms of the sample covariance matrix.
We used the implementation of the $\ell_{1}$-penalized Exact log-likelihood for Matlab made available at Prof. Alexandre D'Aspremont's web site (\texttt{http://www.princeton.edu/$\sim$aspremon/}).

  Boxplots of the different accuracy measures for each of the methods and selection criteria are shown in Figures \ref{figure:estimates_accuracies_star_cases_n_0020}, \ref{figure:estimates_accuracies_arlike_cases_n_0020} and \ref{figure:estimates_accuracies_random_cases_n_0020} for the small sample case ($n=20$), and in Figures \ref{figure:estimates_accuracies_star_cases_n_1000}, \ref{figure:estimates_accuracies_arlike_cases_n_1000} and \ref{figure:estimates_accuracies_random_cases_n_1000} for the large sample case ($n=1,000$).
For larger sample sizes, the Cholesky estimates do suffer some deterioration in terms of the entropy and quadratic losses when an inappropriate ordering of the variables is used.
As we will later see, an inappropriate ordering can also degrade the Cholesky performance in terms of selecting the right covariance terms.

A comparison of the different methods reveals that the best method to use depends on whether the metric used is on the inverse covariance (precision
matrix) or covariance matrix.

\begin{itemize}
  \item With respect to all
the three metrics on the inverse covariance (quadratic, entropy and spectral norm losses), the best results are achieved by SPLICE. 
In the case of the quadratic loss, this result can be partially attributed to the similarity between the quadratic loss and the pseudo-likelihood function used in the SPLICE estimate. 
In terms of the spectral norm on the precision matrix ($\|\bfC-\hat\bfC\|_{2}$), SPLICE performed particularly well in larger sample sizes ($n=1,000$).
For the quadratic and entropy loss functions, AIC was the criterion picking the best SPLICE estimates. 
In terms of the spectral norm loss, SPLICE performs better when coupled with $\aicc$ for small sample sizes ($n=20$) and when coupled with BIC in larger sample sizes ($n=1,000$).

  \item In terms of the spectral norm of the covariance estimate deviation ($\|\Sigma-\hat\Sigma\|_{2}$), the best performance was achieved by Exact PML.
The performance of Exact PML was somewhat insensitive to the selection criterion used in many cases: this may be caused by the uniform grid on $\lambda$ missing regions where the penalized models rapidly change as $\lambda$ varies.
Based on the cases where the selection criterion affected the performance of Exact PML, BIC seems to yield the best results in terms of $\|\Sigma-\hat\Sigma\|_{2}$.

\end{itemize}

For ease of reference, these results are collected in Table \ref{table:suitable_methods_for_different_cases}.

\begin{table}[t]
\begin{center}
\begin{tabular}{c|c}
  \phantom{MMMMMMMMMMMMMMMMMMM}
  &
  \phantom{MMMMMMMMMMMMMMMMMMM}
  \\
  {\bfseries Performance Metric}
  &
  {\bfseries Recommended procedure} \\
  \\
  \hline
  \hline
  \\
%  Quadratic Loss 
  $\Delta_{2}(\bfC, \hat{\bfC})$
  & 
  SPLICE + AIC
  \\
  \\
  \hline
  \\
%  Entropy Loss  
  $\Delta_{e}(\bfC, \hat{\bfC})$
  &
  \begin{tabular}{c}
    SPLICE + AIC
  \end{tabular}
  \\
  \\
  \hline
  \\
%  Spectral norm of $\hat{\bfC}-\bfC$	
  $\|\hat{\bfC}-\bfC\|_{2}$
  &
  \begin{tabular}{c}
  SPLICE + $\aicc$ (for smaller sample size, $n=20$)
  \\  	
  SPLICE + BIC (for larger sample size, $n=1,000$)
  \end{tabular}
  \\
  \\
  \hline
  \\
%  Spectral norm of $\hat{\Sigma}-\Sigma$
  $\|\hat{\Sigma}-\Sigma\|_{2}$
  &
  Exact penalized ML + AIC	
  \\
  \\
  \hline
\end{tabular}

\caption
{
{\bfseries Suitable estimation methods for different covariance structures and performance metrics:}
\label{table:suitable_methods_for_different_cases}
The results shown here are a summary of the results shown in Figures \ref{figure:estimates_accuracies_star_cases_n_0020} through \ref{figure:estimates_accuracies_random_cases_n_1000}.
For each metric, we show the best combination of estimation method and selection criterion based on our simulations.
}
\end{center}	
\end{table}

%%%%%%%%%%%%%%%%%%%%%%%%%%%%%%%%%%%%%%%%%%%%%%%%%%%%%%%%%%%%%%%%%%%%%%%%%%%%%%%%%%%%%%%%%%%%%%

% Comparison to Cholesky, LW and Exact for AR cases

\renewcommand{\figurewidth}{0.85}
\renewcommand{\raiselegendtext}{0.5cm}
\begin{figure}[p]
\begin{center}
\includegraphics[width=\figurewidth\textwidth]{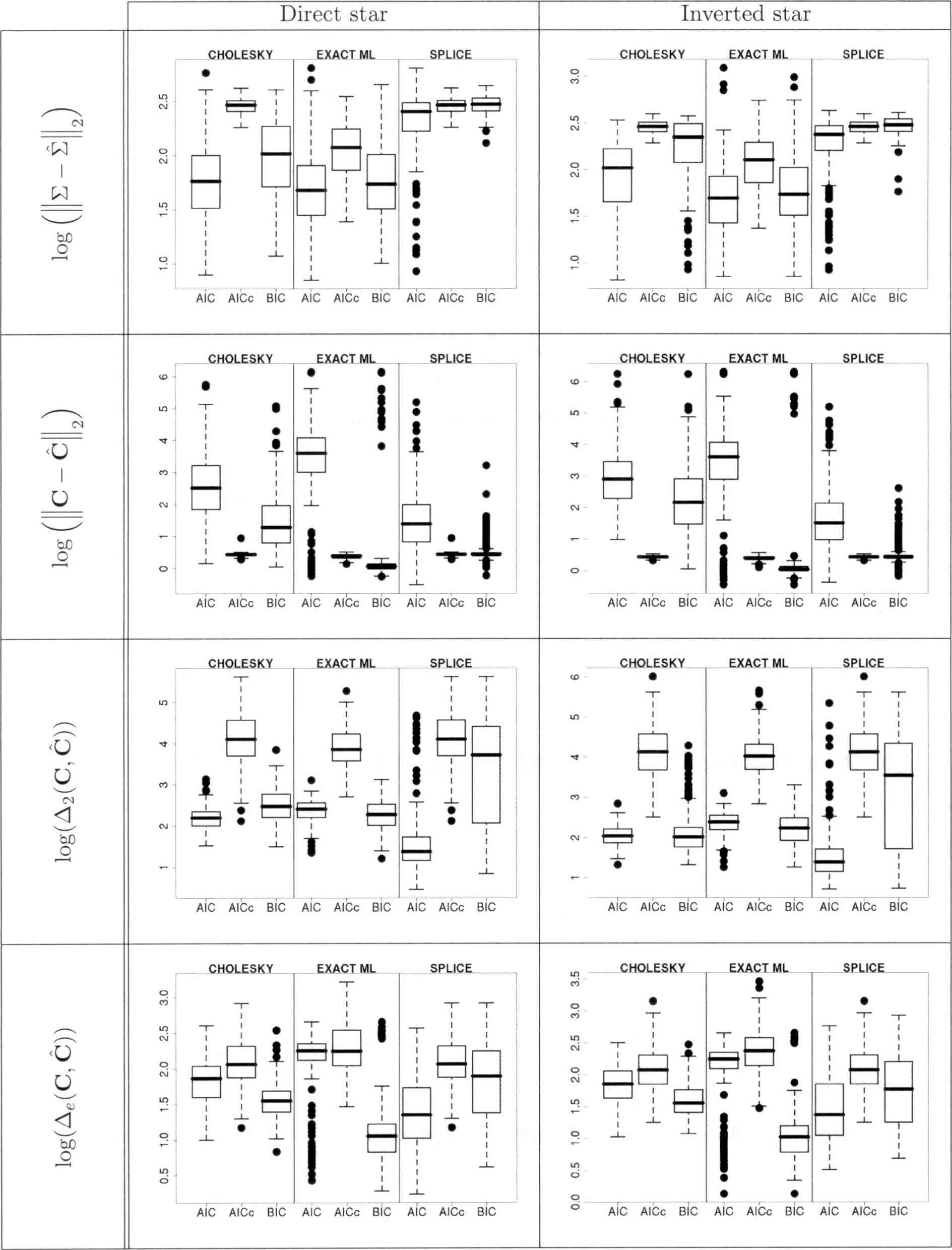}
\caption{
\label{figure:estimates_accuracies_star_cases_n_0020}
{\bfseries Accuracy metrics for precision matrix estimates in the Star cases for $p=15$ and $n=20$}
}
\end{center}
\end{figure}

\renewcommand{\figurewidth}{1.00}
\renewcommand{\raiselegendtext}{0.5cm}
\begin{figure}[p]
\begin{center}
\includegraphics[width=\figurewidth\textwidth]{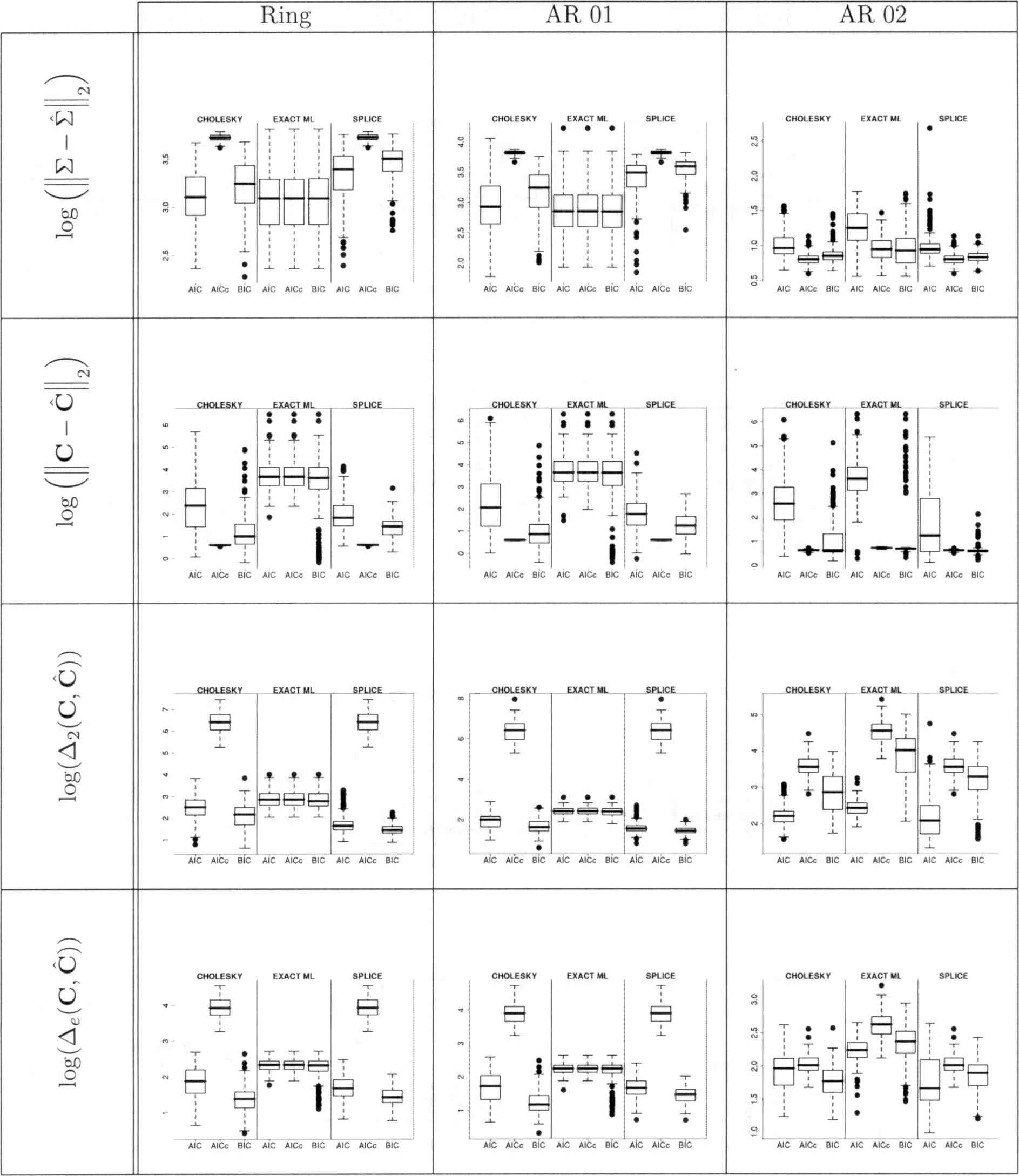}
\caption{
\label{figure:estimates_accuracies_arlike_cases_n_0020}
{\bfseries Accuracy metrics for precision matrix estimates in the AR-like cases for $p=15$ and $n=20$}
}
\end{center}
\end{figure}

\renewcommand{\figurewidth}{1.00}
\renewcommand{\raiselegendtext}{0.5cm}
\begin{figure}[p]
\begin{center}
\includegraphics[width=\figurewidth\textwidth]{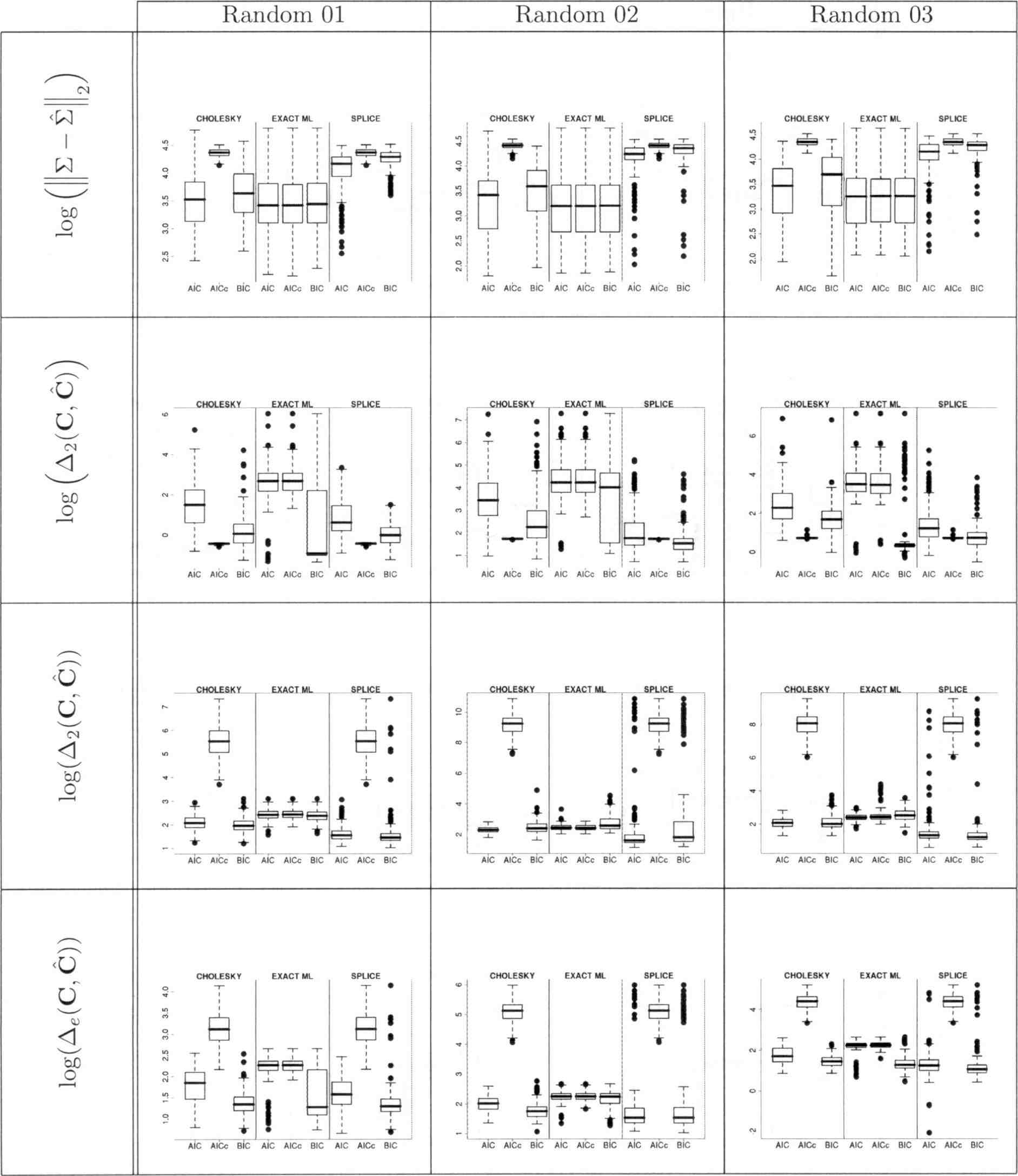}
\caption{
\label{figure:estimates_accuracies_random_cases_n_0020}
{\bfseries Accuracy metrics for precision matrix estimates in the randomly generated designs for $p=15$ and $n=20$}
}
\end{center}
\end{figure}

\renewcommand{\figurewidth}{0.85}
\renewcommand{\raiselegendtext}{0.5cm}
\begin{figure}[p]
\begin{center}
\includegraphics[width=\figurewidth\textwidth]{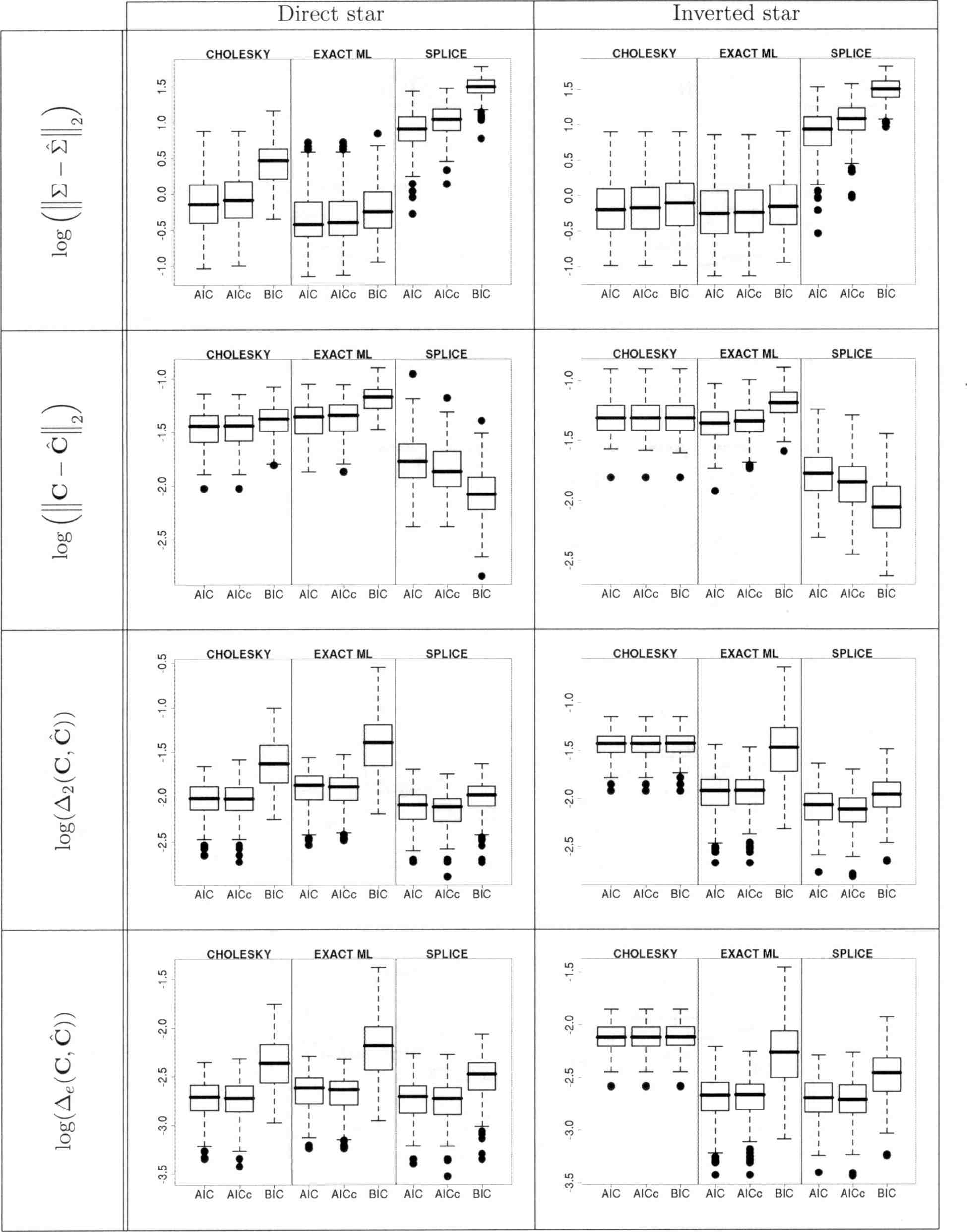}
\caption{
\label{figure:estimates_accuracies_star_cases_n_1000}
{\bfseries Accuracy metrics for precision matrix estimates in the Star cases for $p=15$ and $n=1,000$:}
}
\end{center}
\end{figure}

\renewcommand{\figurewidth}{1.00}
\renewcommand{\raiselegendtext}{0.5cm}
\begin{figure}[p]
\begin{center}
\includegraphics[width=\figurewidth\textwidth]{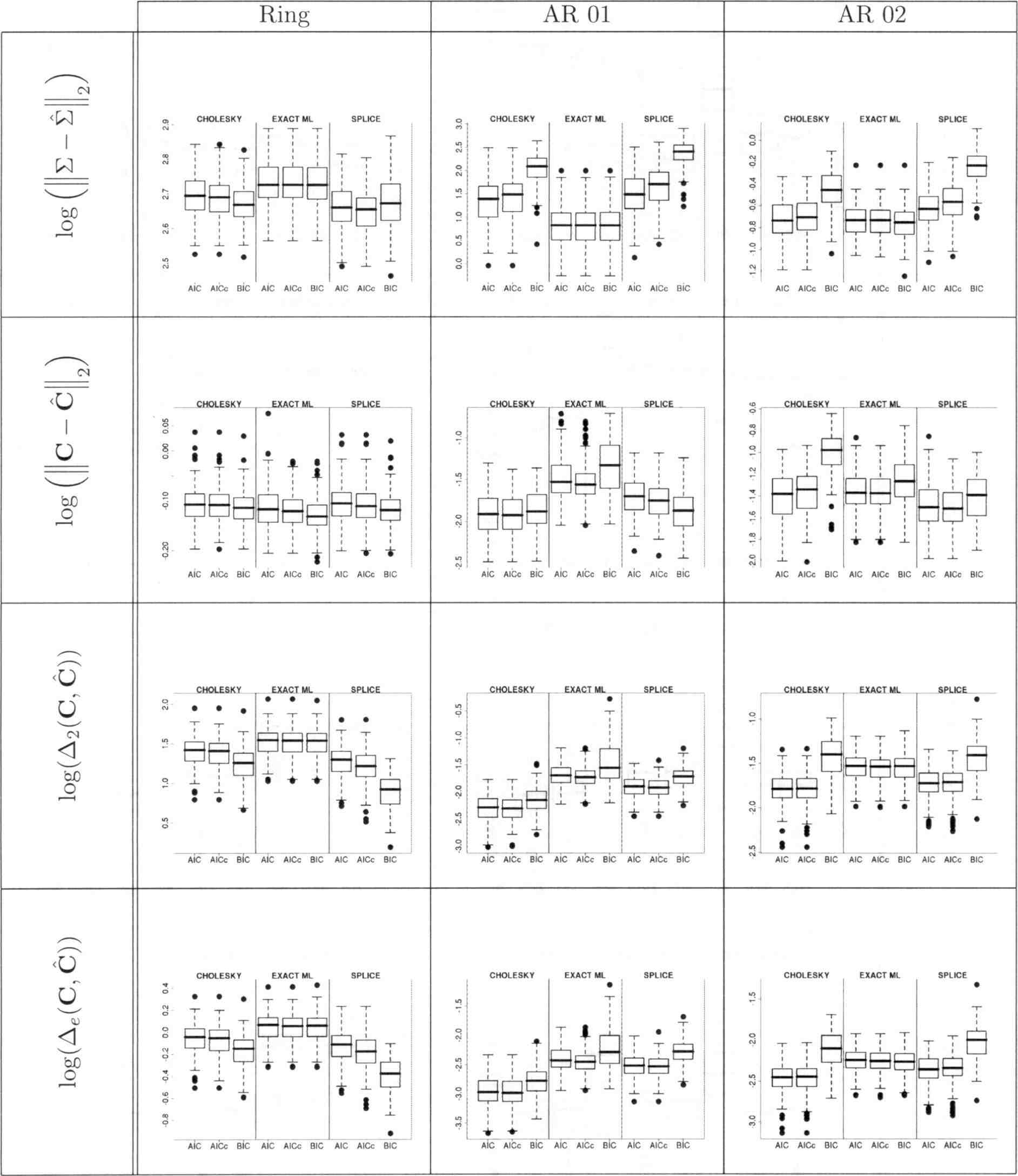}
\caption{
\label{figure:estimates_accuracies_arlike_cases_n_1000}
{\bfseries Accuracy metrics for precision matrix estimates in the AR-like cases for $p=15$ and $n=1,000$}
}
\end{center}
\end{figure}

\renewcommand{\figurewidth}{1.00}
\renewcommand{\raiselegendtext}{0.5cm}
\begin{figure}[p]
\begin{center}
\includegraphics[width=\figurewidth\textwidth]{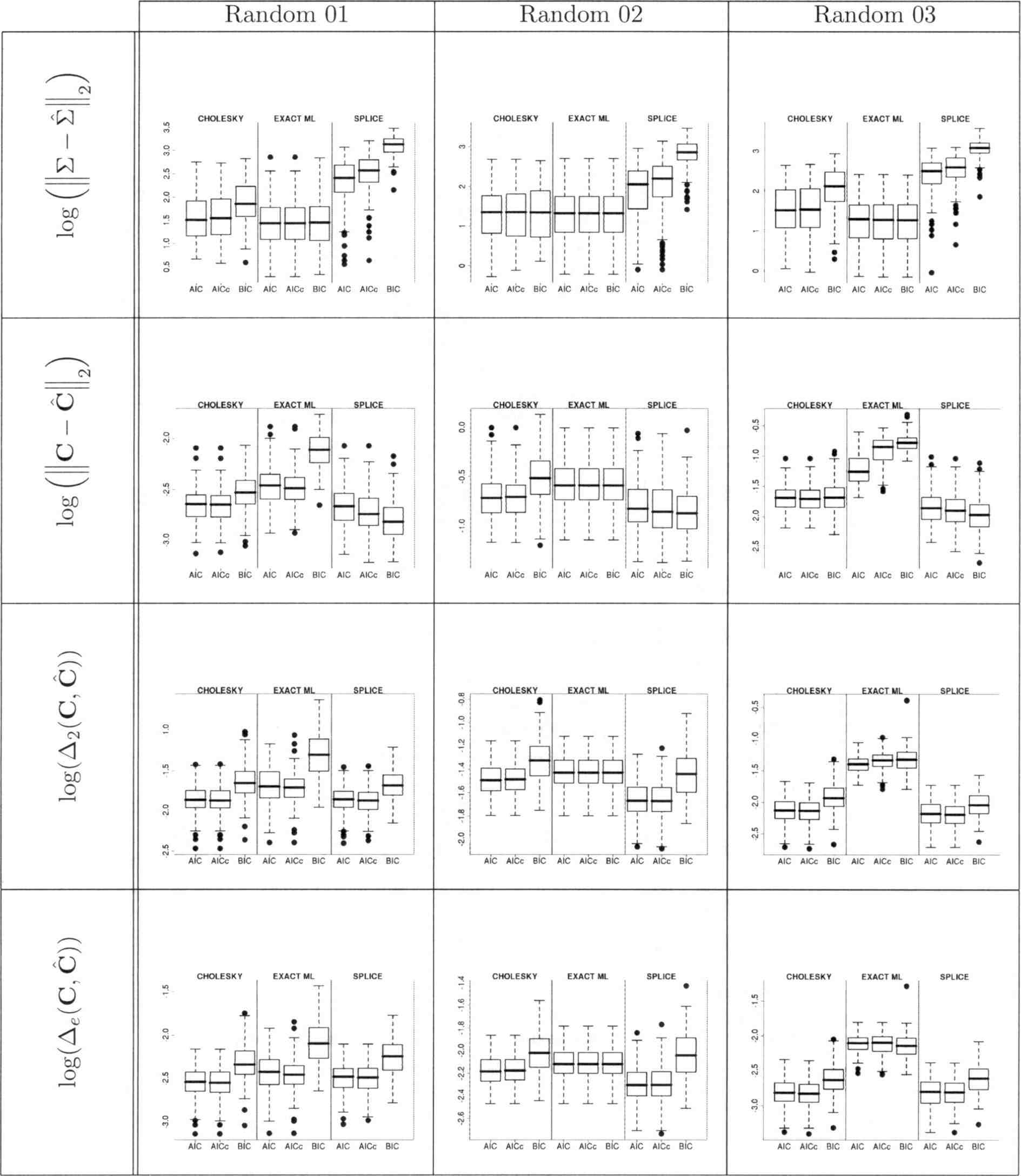}
\caption{
\label{figure:estimates_accuracies_random_cases_n_1000}
{\bfseries Accuracy metrics for precision matrix estimates in the randomly generated designs for $p=15$ and $n=1,000$}
}
\end{center}
\end{figure}

\subsubsection{Model Selection performance of SPLICE}
\label{section:covariance_selection_comparison}

To evaluate the model selection performance of the different covariance selection methods, we compare their Relative Operating Characteristic (ROC) curves defined as a curve containing in its horizontal axis the minimal number of false positives that is incurred on average for a given number of true number of positives, shown on the vertical axis, to be identified.
%In the ROC curves shown in Figures \ref{figure:splice_vs_cholesky_roc_curves_n_0020} and \ref{figure:splice_vs_cholesky_roc_curves_n_1000} the horizontal axis on each panel contains the average of the minimal number of false positives that must be incurred so the 
The ROC curve for a method shows its model selection performance over all possible choices of the tuning parameter $\lambda$.
We have chosen to compare ROC curves instead of the results for particular selection criterion as different applications may penalize false positives and negatives differently.

%These curves thus show the trade-off between the gain of adding a relevant variable and the loss of including an irrelevant variable as we move along the regularization path.
For a fixed number of true positives on the vertical axis, we want the expected minimal number of false positives to be as low as possible.
A covariance selection method is thus better the closer its ROC curve is to the upper left side of the plot.

\renewcommand{\figurewidth}{0.70}
\renewcommand{\raiselegendtext}{0.5cm}
\begin{sidewaysfigure}[p]
\begin{center}
\includegraphics[width=\figurewidth\textwidth]{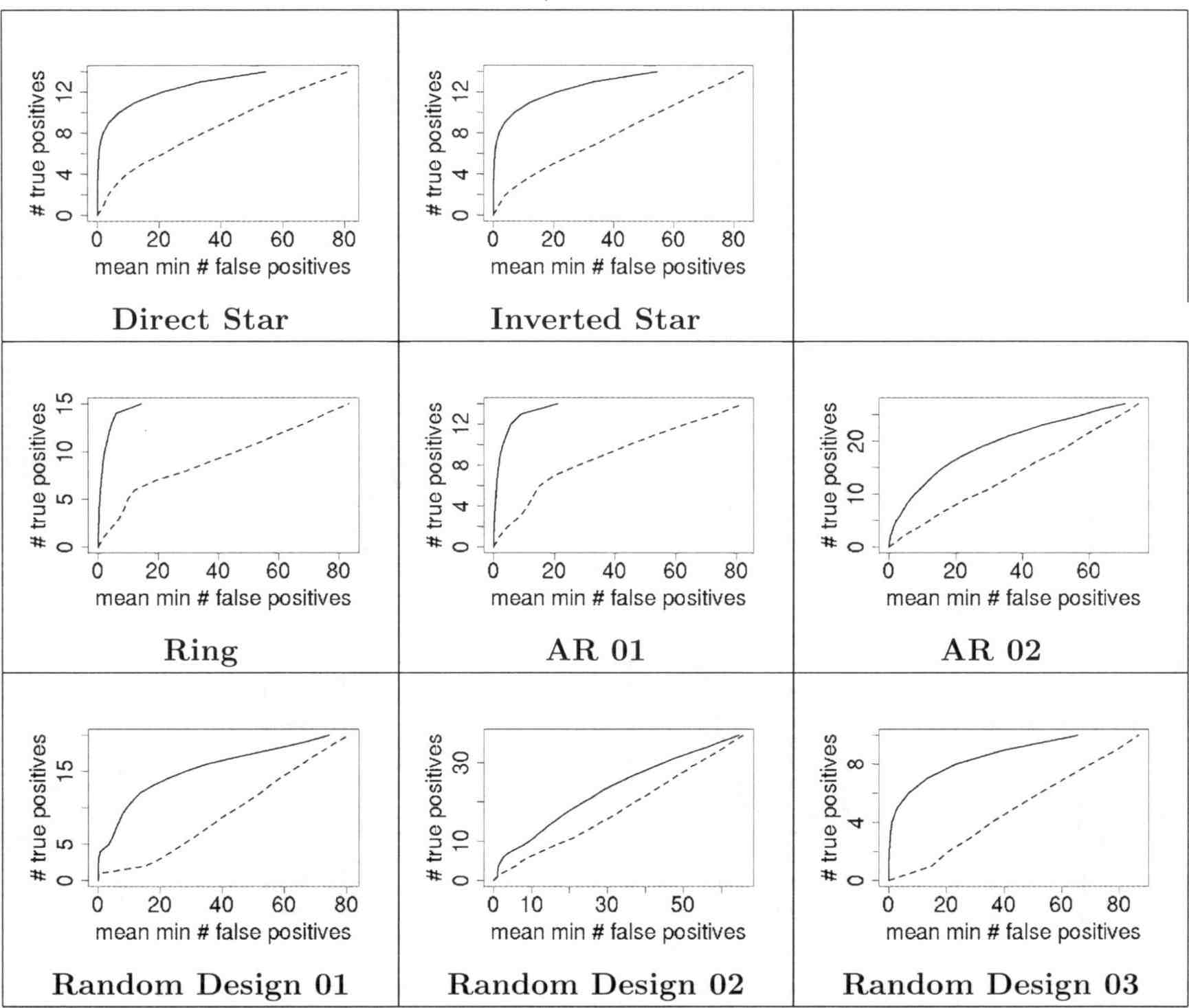}
\end{center}
\begin{small}
\caption
{
{\bfseries ``Mean ROC curves'' for the Cholesky and SPLICE covariance selection for $p=15$ and $n=20$:}
\label{figure:splice_vs_cholesky_roc_curves_n_0020}
Within each panel, the  relative operating characteristic (ROC) curve shows 
the mean minimal number of false positives (horizontal axis) needed to achieve a given number of true positives (vertical axis)
for both the SPLICE (solid lines) and Cholesky (dashed lines).
A selection procedure is better the more its curve approaches the upper left corner of the plot.
Our results suggest that SPLICE trades off better than Cholesky between false and true positives across all cases considered.}
\end{small}
\end{sidewaysfigure}

\renewcommand{\figurewidth}{0.70}
\renewcommand{\raiselegendtext}{0.5cm}
\begin{sidewaysfigure}[p]
\begin{center}
\includegraphics[width=\figurewidth\textwidth]{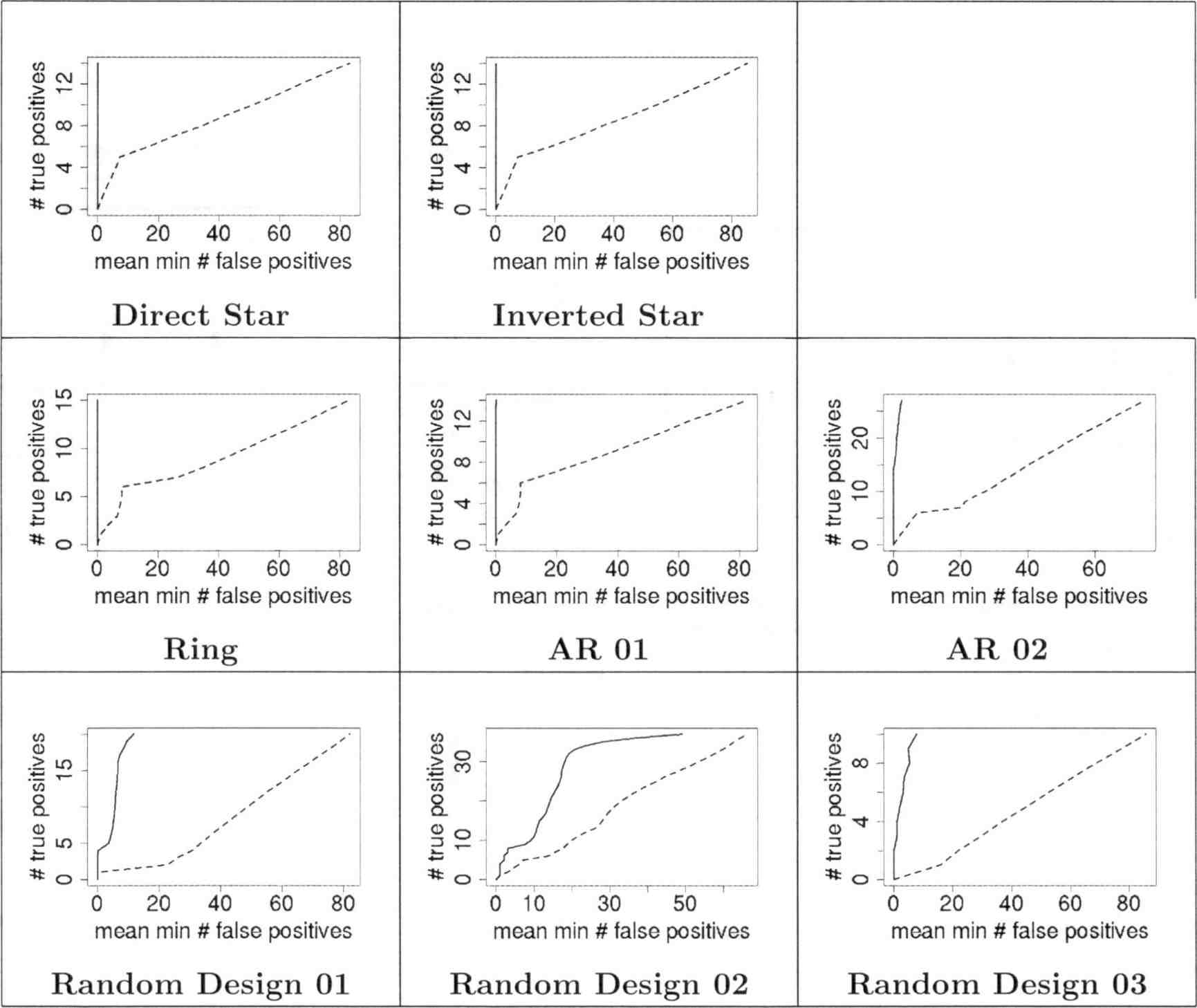}
\end{center}
\begin{small}
\caption
{
{\bfseries ``Mean ROC curves'' for the Cholesky and SPLICE covariance selection for $p=15$ and $n=1,000$:}
\label{figure:splice_vs_cholesky_roc_curves_n_1000}
Within each panel, the  relative operating characteristic (ROC) curve shows 
the mean minimal number of false positives (horizontal axis) needed to achieve a given number of true positives (vertical axis)
for both the SPLICE (solid lines) and Cholesky (dashed lines).
A selection procedure is better the more its curve approaches the upper left corner of the plot.
Our results suggest that SPLICE picks correct covariance terms with a larger sample size for large enough sample sizes.
The Cholesky estimates, however, do not seem much more likely to select correct covariance terms in this larger sample size in the comparison with the case $n=20$.}
\end{small}
\end{sidewaysfigure}

Figures \ref{figure:splice_vs_cholesky_roc_curves_n_0020} and \ref{figure:splice_vs_cholesky_roc_curves_n_1000} compare the ROC curves for the Cholesky and SPLICE covariance selection procedures for sample sizes $n=20$ and $n=1,000$ respectively.
The Exact ML does not have its ROC curve shown: the grid used to approximate its regularization path often did not include estimates with a specific number of true positives.
A finer grid can ameliorate the problem, but would be prohibitively expensive to compute (recall we used a grid with 500 equally spaced values of $\lambda$).
This illustrates an advantage of path following algorithms over using grids: path following performs a more thorough search on the space of models.

The mean ROC curves on Figures \ref{figure:splice_vs_cholesky_roc_curves_n_0020}  and \ref{figure:splice_vs_cholesky_roc_curves_n_1000} show that SPLICE had a better performance in terms of model selection in comparison to the Cholesky method over all cases considered. 
In addition, for a given number of true positives, the number of false positives incurred by SPLICE decreases significantly as the sample size increases.
Our results also suggest that, with the exception of the Random Design 02, the chance that the SPLICE path contains the true model approaches one as the sample size increases for all simulated cases.

Finally, we consider the effect of ordering the variables on the selection performance of SPLICE and Cholesky.
To do this, we restrict attention to the ``star'' cases and compare the performance of SPLICE and Cholesky when the variables are presented in the ``correct'' order ($X_{1}, X_{2}, \ldots, X_{p}$) and in the inverted order ($X_{p}, X_{p-1}, \ldots, X_{1}$).
Figure \ref{figure:splice_vs_cholesky_roc_rows} shows the boxplot of the minimal number of false positives on 200 replications of the Cholesky and SPLICE paths for selected number of true positives in the small sample case ($n=20$).
In addition to outperforming Cholesky by a wide margin, SPLICE is not sensitive to the order in which the variables are presented.
Cholesky, on the other hand, suffers some further degradation in terms of model selection when the variables are presented in the reverse order.

\renewcommand{\figurewidth}{1.00}
\renewcommand{\raiselegendtext}{0.5cm}
\begin{sidewaysfigure}[p]
\begin{center}
\includegraphics[width=\figurewidth\textwidth]{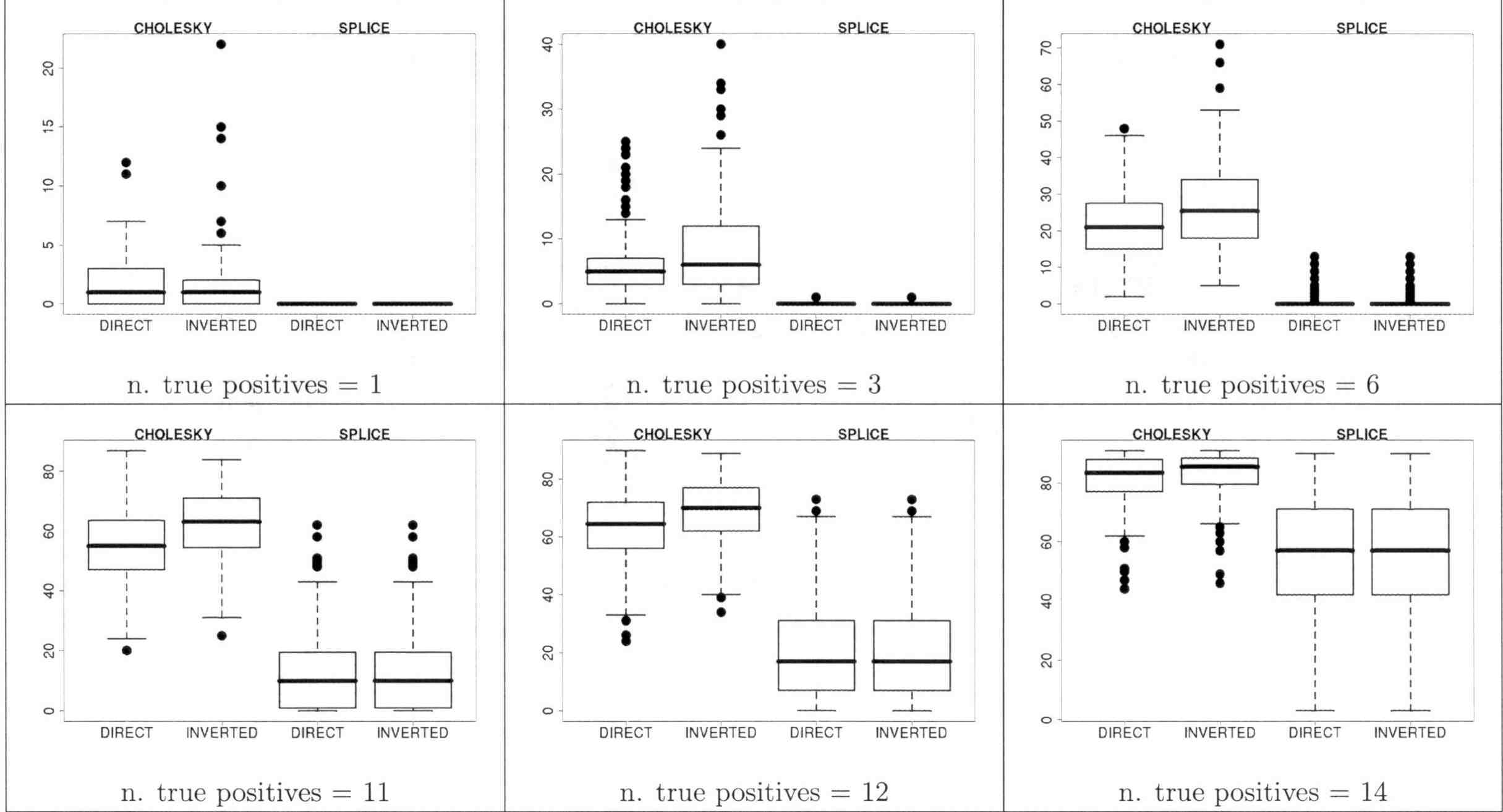}
\end{center}
\begin{small}
\caption
{
{\bfseries Horizontal detail for ROC curves in the ``star'' cases:}
\label{figure:splice_vs_cholesky_roc_rows}
Within each panel, we show a boxplot of the minimal number of false positives on the path for the indicated number of true positives.
Each panel can be thought of as the boxplot corresponding to a horizontal layer on the graph shown in Figure \ref{figure:splice_vs_cholesky_roc_curves_n_0020}.
SPLICE is insensitive to the permutation of the variables (compate inverted and direct).
Cholesky performs worse than SPLICE in both the direct and inverted cases: its performance deteriorates further if the variables are presented in inverted order.
}
\end{small}
\end{sidewaysfigure}

%%%%%%%%%%%%%%%%%%%%%%%%%%%%%%%%%%%%%%%%%%%%%%%%%%%%%%%%%%%%%%%%%%%%%%%%%%%%%%%%%%%%%%%%%%%%%%

\subsection{Positive Semi-Definiteness along the regularization path}
\label{section:psd_experiment}

%As noted in Section \ref{} above, we are unable to prove that the SPLICE estimates are positive semi-definite through its entire regularization path.
%However, we have also pointed out that, at a neighborhood of the beginning of the path, the SPLICE estimates are bound to be positive semi-definite.
As noted in Section \ref{section:pseudolikelihood_loss_function} above, there is no theoretical guarantee that the SPLICE estimates be positive semi-definite.
In the somewhat well-behaved cases studied in the experiments of Section \ref{section:simulation_results_comparison} (see Figure \ref{figure:simulated_topologies}) all of the estimates selected by either $\aicc$ and BIC were positive semi-definite cases.
In only 6 out of the 1,600 simulated cases, did AIC choose a slightly negative SPLICE estimate.
This, however, tells little about the positive-definiteness of SPLICE estimates in badly behaved cases.
We now provide some experimental evidence that the SPLICE estimates can be positive semi-definite for most of the regularization path even when the true covariance matrix is nearly singular.

The results reported for this experiment are based on $200$ replications of SPLICE applied to a data matrix $\bfX$ sampled from a Gaussian distribution with near singular covariance matrix.
The number of observations ($n=40$) and the dimension of the problem ($p=30$) are kept fixed throughout.
To obtain a variety of near singular covariance matrices, the sample covariance $\Sigma\in\bbR^{p\times p}$ of each of the $200$ replications is sampled from:
\begin{eqnarray*}
\Sigma & \sim & \mbox{Wishart}(n, \bar{\Sigma}), \mbox{ with } [\bar{\Sigma}]_{ij} = 0.99^{|i-j|}.
\end{eqnarray*}
The covariance matrices sampled from distribution have expected value $\bar{\Sigma}$, which is itself close to singular.
We let the number of degrees of freedom of the Wishart distribution be small (equal to the sample size $n=30$) to make designs close to singular more likely to happen.
Once $\Sigma$ is sampled, the data matrix $\bfX$ is then formed from an i.i.d. sample of a $N(0,\Sigma)$ distribution.

To align the results along the path of different replications, we create an index $\bar{\lambda}$ formed by dividing a $\lambda$ on the path by the the maximum $\lambda$ at that path.
This index varies over $[0,1]$ and lower values of $\bar\lambda$ correspond to less regularized estimates.
Figure \ref{figure:splice_minimum_eigenvalue_path} shows the minimum eigenvalue of $\hat{\bfC}(\lambda)$ versus the index $\bar\lambda$ for the $200$ simulated replications. 
We can see that non-positive estimates only occur near the very end of the path (small values of $\lambda$) even in such an extreme design. 

\renewcommand{\figurewidth}{0.45}
\afterpage{
\clearpage
\begin{sidewaysfigure}
\begin{center}
\begin{tabular}{cc}
  \includegraphics[width=\figurewidth\textwidth]
  {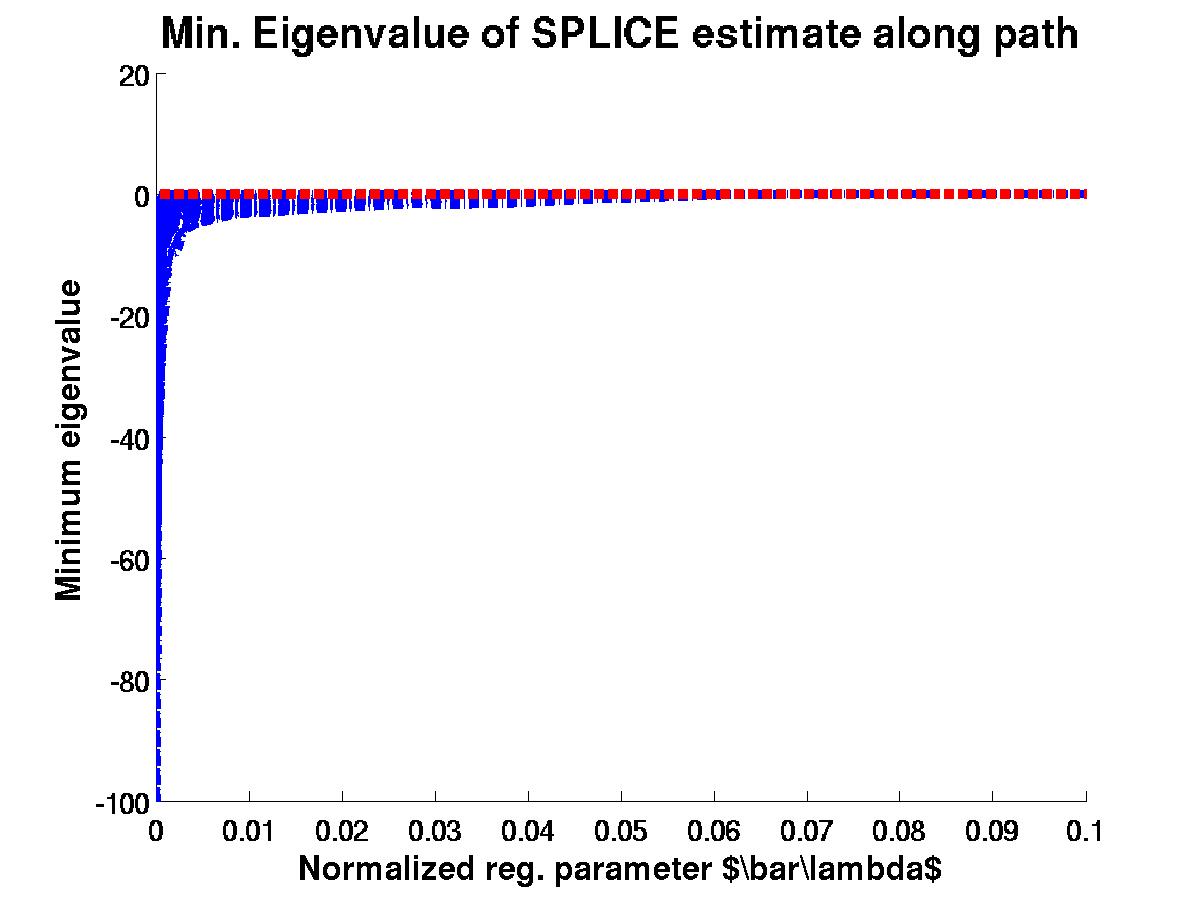}
  &
  \includegraphics[width=\figurewidth\textwidth]
  {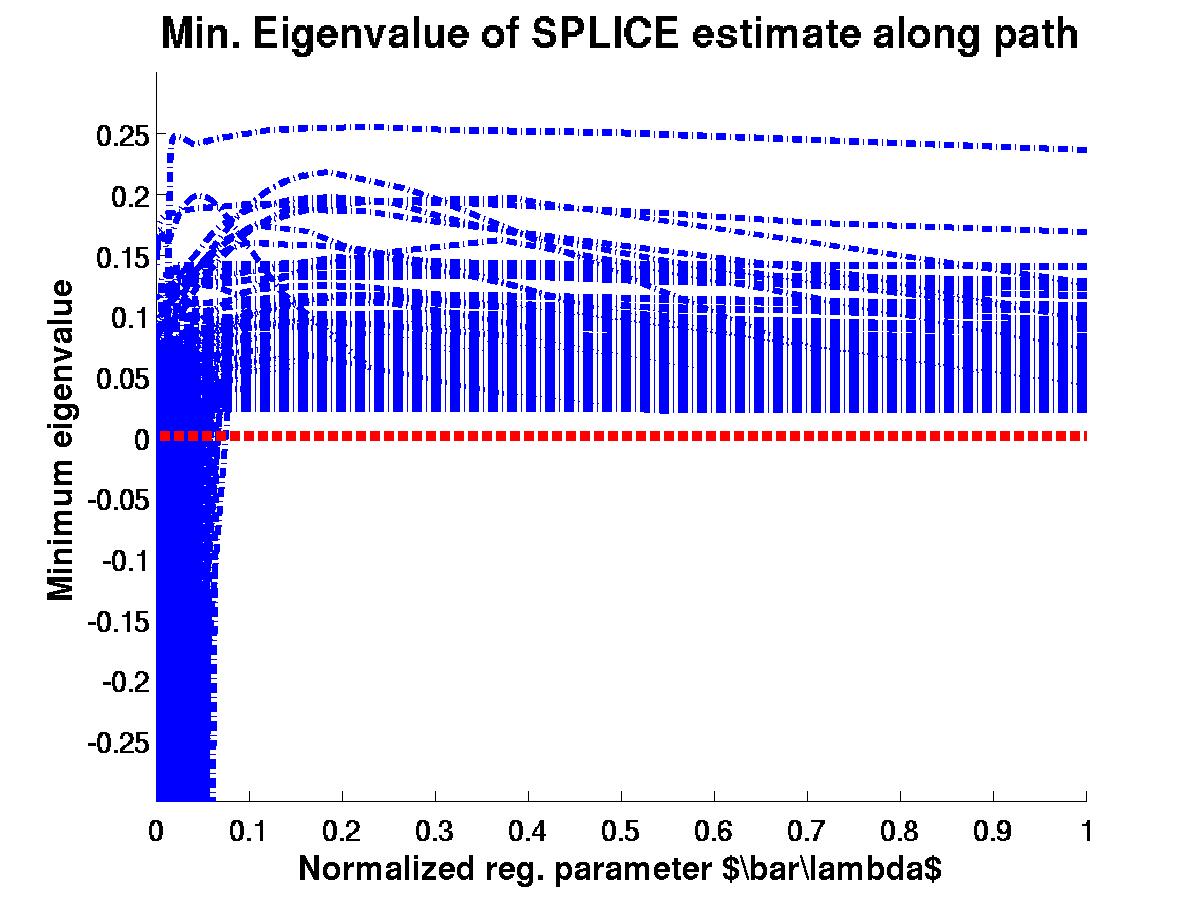}
\end{tabular}
\end{center}	
\begin{small}
\caption
[Positive definiteness along SPLICE path]
{
\label{figure:splice_minimum_eigenvalue_path}
\textbf{Positive definiteness along SPLICE path:}
At different vertical and horizontal scales, the two panels show the the minimum eigenvalue of the SPLICE estimate $\hat{\bfC}(\lambda)$ along the regularization path as a function of the normalized regularization parameter $\bar{\lambda}$.
At the unregularized extreme (small $\bar\lambda$), $\hat\bfC$ can have very negative eigenvalues (left panel).
However, for a long stretch of the regularization path -- over 90\% of the path as measured by the index $\lambda$ -- the SPLICE estimate $\hat\bfC$ is positive definite (right panel).
}
\end{small}
\end{sidewaysfigure}
\clearpage
}

%%%%%%%%%%%%%%%%%%%%%%%%%%%%%%%%%%%%%%%%%%%%%%%%%%%%%%%%%%%%%%%%%%%%%%%%%%%%%%%%%%%%%%%%%%%%%%

%\section{Application to vibration data from Golden Gate Bridges}
%\label{section:goldengate_data_results}
%
%[TODO: This section is not started yet]

%%%%%%%%%%%%%%%%%%%%%%%%%%%%%%%%%%%%%%%%%%%%%%%%%%%%%%%%%%%%%%%%%%%%%%%%%%%%%%%%%%%%%%%%%%%%%%

\section{Discussion}
\label{section:discussion}

In this paper, we have defined Sparse Pseudo-Likelihood Inverse Covariance Estimates (SPLICEs) as a $\norml{1}$-penalized pseudo-likelihood estimate for precision matrices. 
The SPLICE loss function \eqref{equation:gaussian_joint_pseudo_loglikelihood} is obtained from extending previous work by \citet{meinshausen:2006:high-dimensional-graphs-and-variable-selection-with-the-lasso} to obtain estimates of precision matrix that are symmetric.
The SPLICE estimates are formed from estimates of the coefficients and  variance of the residuals of linear regression models.

The main advantage of the estimates proposed here is algorithmic.
The regularization path for SPLICE estimates can be efficiently computed by alternating the estimation of coefficients and variance of the residuals of linear regressions.
For fixed estimates of the variance of residuals, the complete path of the regression coefficients can be traced efficiently using an adaptation of the homotopy/LARS-LASSO algorithm \citep{osborne:2000:lasso_dual, efron:2004:lars} that enforces the symmetry constraints along the path.
Given the path of regression coefficients, the variance of the residuals can be estimated by means of closed form solutions.
An analysis of the complexity of the algorithm suggests that early stopping can reduce its computational cost further.
A comparison of the pseudo-likelihood approximation to the exact likelihood function provides another argument in favor of early stopping:
the pseudo-likelihood approximation is better the sparser the estimated model. 
Thus moving on to the lesser sparse stretches of the regularization path can be not only computationally costly but also counterproductive to the quality of the estimates.

We have compared SPLICE with $\ell_{1}$-penalized covariance estimates based on Cholesky decomposition \citep{huang:2006:covariance-matrix-selection-and-estimation-via-penalized-normal-likelihood} and the exact likelihood expression \citep{banerjee:2005:sparse-covariance-selection-via-robust-maximum-likelihood-estimation, banerjee:2007:model-selection-through-sparse-maximum-likelihood-estimation-for-multivariate-gaussian-or-binary, yuan:2007:model-selection-and-estimation-in-the-gaussian-graphical-model, friedman:2008:sparse-inverse-covariance-estimation-with-the-graphical-lasso} for a variety of sparse precision matrix cases and in terms of four different metrics, namely: quadratic loss, entropy loss, spectral norm of $\bfC-\hat{\bfC}$ and spectral norm of $\Sigma-\hat{\Sigma}$.
SPLICE estimates had the best performance in all metrics with the exception of the spectral norm of $\Sigma-\hat{\Sigma}$.
For this last metric, the best results were achieved by using the $\ell_{1}$-penalized exact likelihood.

In terms of selecting the right terms of the precision matrix, SPLICE was able to pick a given number of correct covariance terms while incurring in less false positives than the $\norml{1}$-penalized Cholesky estimates of \citet{huang:2006:covariance-matrix-selection-and-estimation-via-penalized-normal-likelihood} in various simulated cases.
Using an uniform grid for the exact penalized maximum likelihood provided few estimates with a mid-range number of correctly picked covariance terms.
This reveals that path following methods perform a more thorough exploration of the model space than penalized estimates computed on (pre-determined) grids of values for the regularization parameter.

While SPLICE estimates are not guaranteed to be positive semi-definite along the entire regularization path, they have been observed to be such for most of the path even in a almost singular problem.
Over tamer cases, the estimates selected by AIC, BIC and $\aicc$ were positive semi-definite in the overwhelming majority of cases (1,594 out of 1,600).

%%%%%%%%%%%%%%%%%%%%%%%%%%%%%%%%%%%%%%%%%%%%%%%%%%%%%%%%%%%%%%%%%%%%%%%

\renewcommand{\thesection}{\Alph{section}}
\setcounter{section}{0}

%%%%%%%%%%%%%%%%%%%%%%%%%%%%%%%%%%%%%%%%%%%%%%%%%%%%%%%%%%%%%%%%%%%%%%%

\section{Proofs}
\label{appendix:splice_proofs}

\subsection{Positive Semi-definiteness of $\norml{2}$-penalized Pseudo-likelihood estimate}

We now prove Theorem \ref{theorem:ridged_pseudolikelihood_estimate_is_psd}.
First, we rewrite the $\norml{2}$-norm penalty in a more convenient form:
\begin{eqnarray*}
p+\trace\left[\tilde\bfB'\tilde\bfB\right]
=
p+\sum_{j=1}^{p}\sum_{k=1}^{p}(-b_{jk})^{2}
=
\sum_{j=1}^{p}(1-b_{jj})^{2}+
\sum_{j=1}^{p}\sum_{k=1}^{p}(-b_{jk})^{2}
=
\trace\left[(\bfI_{p}-\tilde\bfB)'(\bfI_{p}-\tilde\bfB)\right]
\end{eqnarray*}

Hence, the $\norml{2}$-penalized estimate defined in \eqref{equation:ridged_pseudo_loglikelihood_estimate}, can be rewritten as:
\begin{eqnarray}
\begin{array}{cccl}
	\hat{\tilde{\bfB}}_{2}(\lambda_{2})
	& = &
	\arg\min\limits_{\tilde{\bfB}} & 
	\mbox{tr}\left[(\bfI_{p}-\tilde{\bfB})\left(\tilde{\bfX}'\tilde{\bfX}+\lambda_{2}\bfI_{p}\right)(\bfI_{p}-\tilde{\bfB}')\right]
	\\
	& & \mbox{s.t.}    &
	 					\left\{
	                    \begin{array}{llll}
		                \tilde{b}_{jj} & = & 0, \mbox{ for } j = 1,\ldots, p \\
						\tilde{b}_{jk} & = & \tilde{b}_{kj}\mbox{ for } j = 1,\ldots, p, k=j+1, \ldots, p\\
						\end{array}\right.		
\end{array}
\label{equation:simplified_ridged_estimate_definition}
\end{eqnarray}

Using convexity, The Karush-Kuhn-Tucker (KKT) conditions are necessary and sufficient to characterize a solution of problem \eqref{equation:simplified_ridged_estimate_definition}:
\begin{eqnarray}
	\left(\tilde{\bfX}'\tilde{\bfX}+\lambda_{2}\bfI_{p}\right)(\bfI_{p}-\hat{\tilde{\bfB}}_{2}(\lambda_{2}))+\Theta+\Omega = 0,
	\label{equation:kkt_conditions_ridge}
\end{eqnarray}
where $\Theta$ is a diagonal matrix and $\Omega$ is an anti-symmetric matrix.
Given that $\tilde{b}_{jj}=0$ and $\omega_{jj}=0$ (anti-symmetry), it follows that, for $\lambda_{2}>0$:
\begin{eqnarray}
	\theta_{jj} & = & -\left(\tilde{\bfX}_{j}'\tilde{\bfX}_{j}+\lambda_{2}\right) < 0,
\end{eqnarray}
that is, $-\Theta$ is a positive definite matrix.

From \eqref{equation:kkt_conditions_ridge}, we can conclude that $\left(\bfI_{p}-\hat{\tilde{\bfB}}_{2}\right)$ satisfies:
\begin{eqnarray*}
	\left(\tilde{\bfX}'\tilde{\bfX}+\lambda_{2}\bfI_{p}\right)\left(\bfI_{p}-\hat{\tilde{\bfB}}_{2}(\lambda_{2})\right)
	+
	\left(\bfI_{p}-\hat{\tilde{\bfB}}_{2}(\lambda_{2})\right)'\left(\tilde{\bfX}'\tilde{\bfX}+\lambda_{2}\bfI_{p}\right)
	& = & 
	-2\Theta.
\label{equation:symetric_kkt_conditions_ridge}
\end{eqnarray*}
Theorem \ref{theorem:ridged_pseudolikelihood_estimate_is_psd} then follows from setting $\bfU=(\bfI_{p}-\hat{\tilde
\bfB}_{2}(\lambda_{2}))$, $\bfV=(\bfX'\bfX+\lambda_{2}\bfI_{p})$ and $\bfW=-\Theta$ and applying the following lemma:
\begin{lemma}
\label{lemma:ridged_estimate_psd}
Let $\bfU$, $\bfV$ and $\bfW$ be $p\times p$ symmetric matrices. Suppose that $\bfV$ is strictly positive definite and $\bfW$ is positive semi-definite and:
\begin{eqnarray*}
\bfU\bfV+\bfV\bfU=\bfW.
\end{eqnarray*}
It follows that $\bfU$ is positive semi-definite.
\end{lemma}

\begin{proof}
	Since $\bfV$ is symmetric positive semi-definite, we can write it as $\bfV=\bfA\Lambda \bfA'$.
	Take a vector $z\in \bbR^{p}$ and rewrite it as $z = \sum_{k=1}^{p}\gamma_{k}a_{k}$ where $a_{k}$ are eigenvectors of $V$ (no need for uniqueness).
	From positive semi-definiteness of $\bfW$ and the assumed identity:
	\begin{eqnarray*}
	0 \le 
	z'\bfW z 
	& = &
	z'\bfU\bfV z + z'\bfV\bfU z 
	\\
	& = & 
	\sum_{j=1}^{p}\sum_{k=1}^{p}\gamma_{j}\gamma_{k}\left(a_{k}'\bfU\bfV a_{j} + a_{k}'\bfV\bfU a_{j}\right) 
	\\
	& = & 
	2\sum_{j=1}^{p}\gamma_{j}^{2}a_{k}'\bfU\bfV a_{k}
	\end{eqnarray*}
	where the second follows from the cross products being zero:
	\begin{eqnarray*}
	a_{j}'\bfU \bfV a_{k} = \mbox{trace}(a_{j}'\bfU\bfV a_{k}) = \mbox{trace}(a_{k}a_{j}'\bfU\bfV) = 0, \mbox{ whenever } j\neq k.
	\end{eqnarray*}

	Now, taking in particular $z=a_{k}$ we have, for every $k=1, \ldots, p$:
	\begin{eqnarray*}
	2a_{k}'\bfU\bfV a_{k} = 
	2\lambda_{k}a_{k}'\bfU a_{k} =
	a_{k}'\bfW a_{k} \ge 0
	\end{eqnarray*}
	We can conclude that for every $k$ having $\lambda_{k}>0$, $a_{k}'\bfU a_{k}\ge 0$ and the result follows.	
\end{proof}

\section{Algorithms}

\subsection{Appendix: A Path-Following Algorithm for the Cholesky estimate}
\label{appendix:cholesky_path_tracing}

In this section, we describe a path tracing algorithm for the precision matrix estimate based on Cholesky decomposition introduced in \citet{huang:2006:covariance-matrix-selection-and-estimation-via-penalized-normal-likelihood}.
The algorithm can be understood as a block-wise coordinate optimization in the same spirit as \citet{friedman:2007:pathwise-coordinate-optimization}.

For a fixed diagonal matrix $\bfD^{2} = \diag\left(d_{1}^{2}, \ldots, d_{p}^{2}\right)$, the sparse Cholesky estimate of \citet{huang:2006:covariance-matrix-selection-and-estimation-via-penalized-normal-likelihood} is:
\begin{eqnarray*}
\begin{array}{lclll}
	\hat{\bfU}(\lambda) & = & \arg\min\limits_{\bfU \in \mbox{UUT}} \bfX\bfU \bfD^{-2} \bfU'\bfX' + \lambda \|\bfU\|_{1},
\end{array}
\end{eqnarray*}
where UUT denotes the space of upper triangular matrices with unit diagonal.
This is equivalent to solving:
\begin{eqnarray*}
\begin{array}{lclll}
	\hat{\beta}(\lambda) 
	& = & 
	\arg\min\limits_{\beta}\sum_{j=1}^{p}\frac{\|\bfX_{j}-\sum_{k=1}^{j-1}\bfX_{k}\beta_{jk}\|^{2}}{d_{j}^{2}} + \lambda \sum_{j=1}^{p}\sum_{k=1}^{j-1}\left|\beta_{jk}\right|.
\end{array}
\label{equation:cholesky_problem}
\end{eqnarray*}

It is not hard to see that the objective function can be broken into $p-1$ uncoupled smalled components.
As a result, the optimization problem can be separated into $p-1$ smaller problems, that is, $\hat{\beta}(\lambda) = \left(\hat{\beta}_{2}(\lambda), \ldots, \hat{\beta}_{p}(\lambda)\right)$ with:
\begin{eqnarray*}
\begin{array}{lclll}
	\hat{\beta}_{j}(\lambda d_{j}^{2}) 
	& = & 
	\arg\min\limits_{\beta}\|\bfX_{j}-\sum_{k=1}^{j-1}\bfX_{k}\beta_{jk}\|^{2}
	+ 
	\lambda d_{j}^{2} \sum_{k=1}^{j-1}\left|\beta_{jk}\right|.
\end{array}
\label{equation:cholesky_subproblems}
\end{eqnarray*}
Each of these $p-1$ subproblems can have its path regularization traced by means of the homotopy/LARS-LASSO algorithm in \citet{osborne:2000:lasso_dual, efron:2004:lars}. The $\hat{\beta}(\lambda)$ parameter estimate is recovered by $\left(\hat{\beta}_{2}(\lambda d_{2}^{2}), \ldots, \hat{\beta}_{p}(\lambda d_{p}^{2})\right)$. All is needed is a little care in merging the $p-1$ paths together as the scaling of the regularization parameter changes from one subproblem to the next.

An alternative way to understand how the problem can be broken into these smaller pieces stems from the representation of program in \eqref{equation:cholesky_problem} as a linear regression.
A little manipulation can be used to show that \eqref{equation:cholesky_problem} can be represented as a linear regression of $\tilde{\bfY}$ against $\tilde\bfZ$ as below:
\begin{eqnarray*}
\begin{array}{cclcccl}
    \tilde{\bfY} 
	 = 
	\left[
	\begin{array}{c}
		\frac{{\bfX}_{2}}{d_{2}^{2}} \\
		\frac{{\bfX}_{3}}{d_{3}^{2}} \\
		\frac{{\bfX}_{4}}{d_{4}^{2}} \\
		\vdots \\
		\frac{{\bfX}_{p-1}}{d_{p-1}^{2}} \\
		\frac{{\bfX}_{p}}{d_{p}^{2}}
	\end{array}
	\right]%_{np\times 1}
	&
	\mbox{ and, }
	\\
    \tilde{Z}
	 = 
%	\left[
%	\hat{\bfD}_{k-1}^{-1}\otimes \bfI_{n}
%	\right]	
	\left[
	\begin{array}{ccccccccccccc}
	\frac{\tilde{\bfX}_{1}}{d_{2}^{2}} & 
	\bfzero & \bfzero & 
	\bfzero & \bfzero & \bfzero & 
	\cdots & 
	\bfzero & \cdots & \bfzero  &
	\bfzero & \cdots & \bfzero
	\\
	\bfzero & 
	\frac{\tilde{\bfX}_{1}}{d_{3}^{2}} & \frac{\tilde{\bfX}_{2}}{d_{3}^{2}} & 
	\bfzero & \bfzero & \bfzero & 
	\cdots &
	\bfzero & \cdots & \bfzero &
	\bfzero & \cdots & \bfzero
	\\
	\bfzero & 
    \bfzero & \bfzero & 
	\frac{\tilde{\bfX}_{1}}{d_{4}^{2}} & \frac{\tilde{\bfX}_{2}}{d_{4}^{2}} & \frac{\tilde{\bfX}_{3}}{d_{4}^{2}} & 
	\cdots &
	\bfzero & \cdots & \bfzero &
	\bfzero & \cdots & \bfzero
	\\
	\vdots & 
	\vdots & \vdots & 
	\vdots & \vdots & \vdots & 
	\ddots &
	\vdots & \ddots & \vdots &
	\vdots & \ddots & \vdots \\
	\bfzero & 
    \bfzero & \bfzero & 
    \bfzero & \bfzero & \bfzero &
    \cdots  & 
    \frac{\tilde{\bfX}_{1}}{d_{p-1}^{2}} & \cdots & \frac{\tilde{\bfX}_{p-2}}{d_{p-1}^{2}} &
	\bfzero & \cdots & \bfzero \\
	\bfzero & 
    \bfzero & \bfzero & 
    \bfzero & \bfzero & \bfzero &
    \cdots  & 
	\bfzero & \cdots & \bfzero &
    \frac{\tilde{\bfX}_{1}}{d_{p}^{2}} & \cdots & \frac{\tilde{\bfX}_{p-1}}{d_{p}^{2}} 
	\end{array}
	\right]%_{np\times \frac{p(p-1)}{2}}	 
\end{array}
\end{eqnarray*}
The separability of the program \eqref{equation:cholesky_problem} into the subprograms \eqref{equation:cholesky_subproblems} follows from the block diagonal structure of the matrix $\tilde\bfZ'\tilde\bfZ$.
The application of the homotopy/LARS-LASSO algorithm to each of the problems and the subsequent merging of the resulting paths into a single path can be seen as a path version of the coordinate wise algorithms described in \citet{friedman:2007:pathwise-coordinate-optimization}.

\section{Appendix: Sampling sparse precision matrices}
\label{section:sampling_precision_matrices}

In Section \ref{section:simulation_results_splice}, we use three randomly selected precision matrices in the simulation studies presented therein.
These random precision matrices are sampled as follows.
\begin{itemize}
  \item A random sample containing 20 observations of $X$ is sampled from a zero-mean Gaussian distribution with precision matrix containing $2$ along its main diagonal and $1$ on its off-diagonal entries;
  \item A random precision matrix is formed by computing $G = (X^{T}X)^{-1}$;
  \item The number of off-diagonal terms $N$ is sampled from the a geometric distribution with parameter $\lambda = 0.05$ conditioned to be between $1$ and $\frac{15\times 14}{2}=105$;
  \item A new matrix $H$ is formed by setting all off-diagonal of the matrix $G$ are set to zero, except for $N$ randomly selected entries (all entries are equally likely to be picked);
  \item Since $H$ may not be positive definite, the precision matrix is formed by adding $H + \bfI_{15}\cdot\max(0, 0.02-{\varphi}(H))$ where ${\varphi}(H)$ is the smallest eigenvalue of $H$.
\end{itemize}

\setlength{\bibsep}{4pt}

\renewcommand{\bibnumfmt}[1]{[\textbf{#1}]}

\begin{small}
%\begin{singlespacing}
\bibliographystyle{acmtrans}
\bibliography{my_papers}
%\end{singlespacing}
\end{small}

%%%%%%%%%%%%%%%%%%%%%%%%%%%%%%%%%%%%%%%%%%%%%%%%%%%%%%%%%%%%%%%%%%%%%%%%%%%%%%%%%%%%%%%%%%%%%%%%%%%%%%%%%%

\end{document}